\documentclass[10pt,twocolumn,letterpaper]{article}

\usepackage[pagenumbers]{cvpr} 

\definecolor{cvprblue}{rgb}{0.21,0.49,0.74}
\usepackage[pagebackref,breaklinks,colorlinks,allcolors=cvprblue]{hyperref}
\usepackage{booktabs}
\usepackage{xcolor}
\usepackage{colortbl}
\usepackage{adjustbox}
\usepackage{graphicx}
\usepackage{amssymb}
\usepackage{multirow}
\usepackage{bm}

\definecolor{TransferOursBg}{HTML}{EDF4FA}
\newcommand{\transferdir}[1]{\rotatebox[origin=c]{90}{\textbf{#1}}}
\newcommand{\transferbest}[1]{\textbf{#1}}
\newcommand{\transfersecond}[1]{\underline{#1}}

\definecolor{RetrievalRefGray}{HTML}{707070}
\definecolor{RetrievalOursBg}{HTML}{EDF4FA}

\newcommand{\retrbest}[1]{\textbf{#1}}
\newcommand{\retrsecond}[1]{\underline{#1}}
\newcommand{\retrref}[1]{\textcolor{RetrievalRefGray}{\textit{#1}}}

\def\confName{CVPR}
\def\confYear{2027}

\title{Learning Binary Codes with Spiking Neural Networks for Cross-Modal Hashing Retrieval}

\author{
Yukuan Zhang$^{1,2}$
Jiarui Zhao$^{1,2}$
Shangqing Nie$^{1,2}$
Jin Kuang$^{3}$
Dong Liu$^4$
Shengsheng Wang$^{1,2}$\thanks{Corresponding author}\\
\small
$^{1}$College of Computer Science and Technology, Jilin University\\
\small
$^{2}$Key Laboratory of Symbolic Computation and Knowledge Engineering of the Ministry of Education, Jilin University\\
\small
$^{3}$School of Computer Science and Engineering, Central South University\\
\small
$^{4}$Hunan Engineering Research Center of Advanced Embedded Computing and Intelligent Medical Systems, Xiangnan University\\
{\tt\small
\{zyk24, zhaojr24, niesq25\}@mails.jlu.edu.cn
}\\
{\tt\small
gasking@csu.edu.cn, liudong@xnu.edu.cn, wss@jlu.edu.cn
}
}

\begin{document}
\maketitle
\begin{abstract}
    Cross-modal hashing (CMH) uses compact binary codes for efficient image-text retrieval. Most methods learn image-text relationships in a continuous space and binarize the representation only at the end. Semantic learning and bit decisions thus remain largely separate.
    This work addresses two key questions concerning hash-bit generation: \textbf{(1) How to determine hash bits?} We leverage the discrete nature of spiking activity and determine each bit through competition between positive and negative spikes accumulated over time, rather than through a single firing event or subsequent thresholding of continuous representations. \textbf{(2) How to generate such hash bits?} We propose an unsupervised spiking framework for cross-modal hashing, in which image and text jointly contribute to hash-bit generation through temporal interaction. An activation constraint further prevents unsupported decisions caused by silent channel pairs. Extensive experiments on three benchmarks demonstrate the effectiveness of our approach and support temporal spike competition as a design principle for hash-code generation.
    \end{abstract}
\begin{figure}[!ht]
    \centering
    \includegraphics[width=0.8\linewidth]{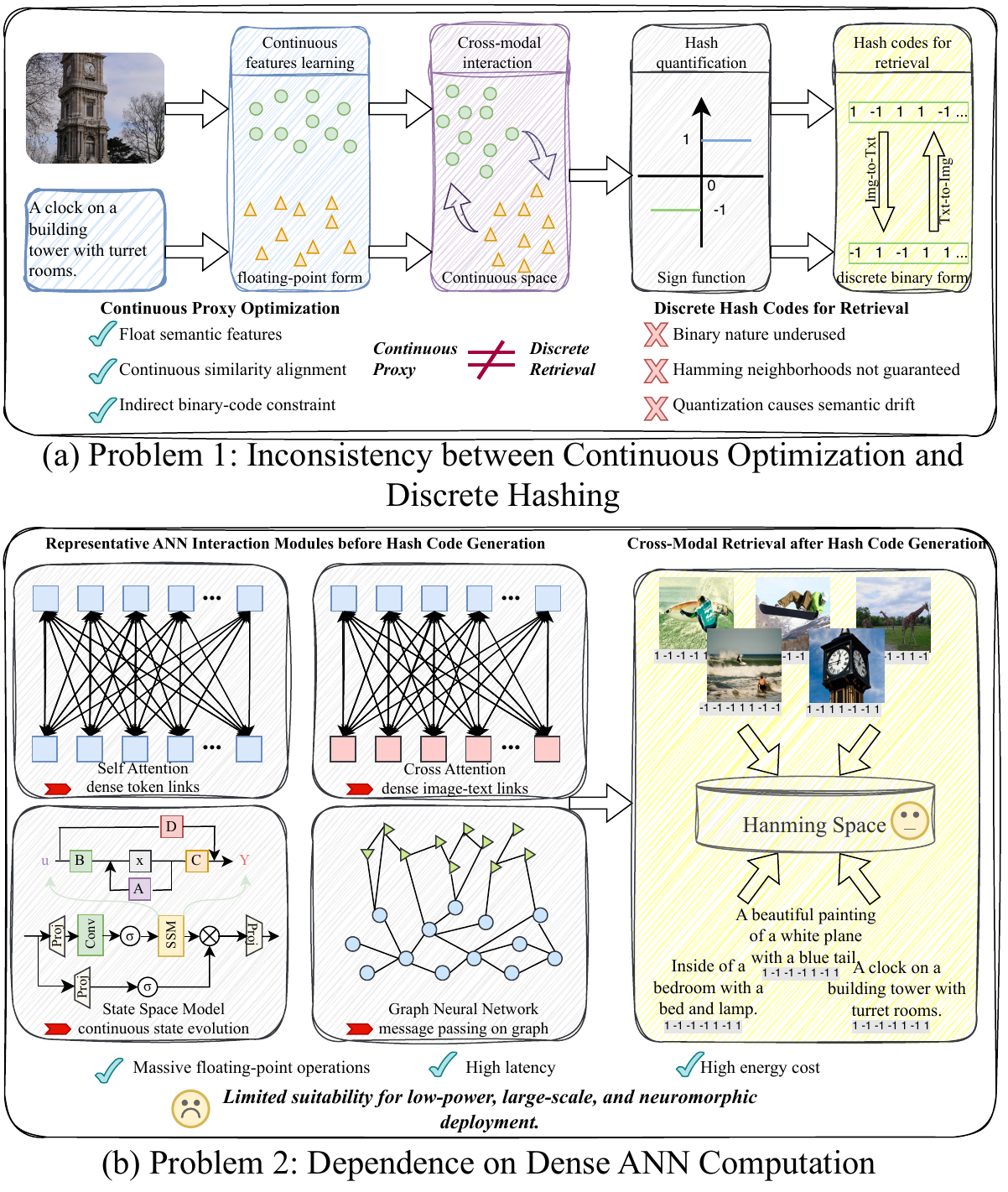} 
    \vspace{-10pt}
    \caption{Typical pipeline and limitations of existing cross-modal hashing methods. }
\label{fig1}
\end{figure}
\section{Introduction}
Retrieval performance in CMH depends on how well binary codes preserve semantic relationships~\cite{hu2024cross, wang2026prior}. Continuous representations capture semantic similarities between images and texts~\cite{li2026joint}, whereas Hamming distance measures differences based solely on bit agreement and disagreement. The shorter the code length, the greater the influence of each bit. Hash-bit generation is therefore not merely a binarization operation but a critical factor in effective cross-modal retrieval.

Modern methods achieve impressive retrieval performance by learning cross-modal semantic relationships in a continuous embedding space~\cite{hu2022unsupervised, sun2026unified}. To this end, they employ neighborhood reconstruction~\cite{su2019deep}, graph-based correlations~\cite{zhang2021aggregation,zhu2022work}, contrastive learning~\cite{cui2024structure,huang2025dual}, clustering~\cite{yang2026stationary,zhao2026udch}, or pretrained vision-language representations~\cite{xia2023clip,huang2025dual}. However, this learning paradigm presents two issues that conflict with the goals of CMH. As shown in Figs.~\ref{fig1}(a) and~\ref{fig1}(b), \textbf{(1) cross-modal semantic learning is separated from the final binary decisions}; and \textbf{(2) hash-bit generation still relies on dense artificial neural network (ANN) computations}, incurring substantial computational and energy costs.

Unifying cross-modal semantic learning with discrete hash-bit generation remains an open challenge. This gap motivates two key questions:
\begin{itemize}
    \item \textbf{\textit{$\bm{Q}_\mathbf{1}$: How to determine hash bits?}}
    \textit{How can reliable bitwise decisions be made in a discrete space without relying on subsequent binarization of continuous representations?}

    \item \textbf{\textit{$\bm{Q}_\mathbf{2}$: How to generate such hash bits?}}
    \textit{Given $\bm{Q}_\mathbf{1}$, how can semantic information from images and texts jointly contribute to hash-bit generation?}
\end{itemize}

To address these questions, we leverage spiking neural networks (SNNs)~\cite{fang2021incorporating} to determine hash bits through temporally accumulated spike activity.

For $\bm{Q}_\mathbf{1}$, a single spike only reflects an instantaneous firing state~\cite{zheng2025spiliformer}. It is insufficient to determine a stable bit value or provide clear support for a binary decision. We therefore determine each bit through the temporal competition of accumulated positive and negative spikes. This allows discrete spiking activity to directly contribute to binary decisions. However, when both channels remain silent, the decision lacks a clear basis. We introduce a bit-activation constraint to discourage such unsupported outcomes. Spikes no longer serve merely as carriers of computation but become the direct basis for hash-bit generation. This insight naturally leads to $\bm{Q}_\mathbf{2}$, \textit{which asks how cross-modal semantics can directly participate in this competitive process.}

For $\bm{Q}_\mathbf{2}$, we argue that the key is to enable images and texts to jointly determine discrete decisions. If cross-modal interaction occurs only in continuous representation learning, binary decisions remain separate from semantic learning. Based on this insight, we propose \textbf{SpikeBit}, which unifies cross-modal semantic interaction and hash-\textbf{bit} generation through a \textbf{spike}-based process. Specifically, both modalities jointly drive the evolution of a shared spiking state and influence each other through bidirectional modulation of firing responses. As a result, cross-modal semantics no longer merely guide continuous features before binarization but directly affect the accumulated competition between positive and negative spikes and the final bit values. Semantic interaction and binary decisions thus become two interconnected parts of the same spiking process.

Our main contributions are as follows:
\begin{itemize}
\item \textbf{A novel discrete hash-bit decision mechanism.} We generate hash bits through temporal competition between positive and negative spikes and employ a bit-activation constraint to reduce unsupported decisions.

\item \textbf{A cross-modal spiking hash-bit generation approach.} We propose SpikeBit, which combines shared spiking state evolution and bidirectional gated interaction to jointly learn cross-modal semantics and generate hash bits.

\item \textbf{Effective retrieval with computational efficiency.} Experiments on three benchmark datasets demonstrate that SpikeBit achieves state-of-the-art or competitive retrieval performance while significantly reducing parameter count, operation count, and estimated energy consumption.
\end{itemize}
\section{Related work}
\noindent\textbf{Cross-modal hashing.} CMH learns compact binary representations in a common space for heterogeneous data~\cite{tu2023unsupervised}. Research spans matrix factorization, binary reconstruction, and latent factor modeling~\cite{ding2014collective,li2017deep,wu2018unsupervised}, neighborhood and graph relation modeling~\cite{su2019deep,zhang2021aggregation,zhu2022work}, structural alignment and semantic decomposition~\cite{cui2024structure,zhang2024unsupervised}, as well as clustering-based discovery, similarity fusion, and Transformer-based modeling~\cite{zhao2026udch,yang2025unsupervised,yang2026stationary}. Recent methods further incorporate CLIP~\cite{radford2021learning} to strengthen cross-modal semantic priors~\cite{xia2023clip,huang2025dual,fan2025vision,tu2022differentiable,tu2025cross,huo2023deep,huo2024deep,tu2024two}. These approaches enrich the sources and modeling of semantic relationships, but semantic consistency in the representation space does not automatically carry over to discrete bit decisions. SpikeBit generates hash bits through competition between positive and negative spikes, applying semantic alignment directly to the responses that determine bit polarity.

\noindent\textbf{Spiking neural networks.}
Spiking neural networks (SNNs) have attracted increasing attention due to their biological plausibility and low-power potential on neuromorphic hardware~\cite{di2025cost,zhang2026safa}. Unlike conventional ANNs, SNNs process information through event-driven neuronal dynamics, including membrane charging, leakage, firing, and resetting. Existing studies have extended spiking computation from convolutional architectures~\cite{li2025brain} to Transformer-style models~\cite{zhou2022spikformer}, and further improved temporal representation through feedback dynamics~\cite{zheng2025stf}, lateral inhibition~\cite{zheng2025spiliformer}, and relative positional encoding~\cite{lv2026toward}.
Beyond basic visual recognition~\cite{yang2026spiketad,xu2026snnsir}, SNNs have recently been applied to more complex tasks, including natural language processing~\cite{lv2025spikebert}, video temporal grounding~\cite{bal2026spikingvtg}, and multimodal pretraining~\cite{lv2025spikeclip}. Spiking signals and hash codes both use binary representations, suggesting a new approach to generating hash bits directly through spiking dynamics. However, existing research~\cite{zong2026brain,zong2026msgat,fu2026brainhash} focuses primarily on the energy efficiency of spiking computation, leaving the connection between spiking dynamics and hash-bit generation underexplored. This work aims to address this gap.
\section{Proposed method}
\subsection{Preliminary definition}
\begin{figure}[!ht]
    \centering
    \includegraphics[width=0.8\linewidth]{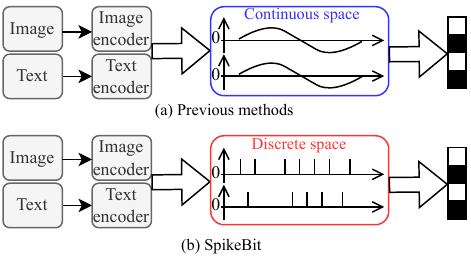} 
    \vspace{-13pt}
    \caption{Comparison of hash-code generation between existing methods and SpikeBit. Both use image and text encoders. Existing methods learn cross-modal relationships in continuous space and generate binary codes at the final stage. In contrast, SpikeBit enables cross-modal interaction through discrete spiking activity, allowing semantic interaction to directly contribute to hash-bit generation.}
    \label{Comparison_of_Framework_Diagrams}
\end{figure}

\noindent\textbf{Cross-modal hashing.} 
Given $N$ paired image-text samples $\mathcal{D}=\{(I_i,T_i)\}_{i=1}^{N}$, we follow existing CMH 
methods~\cite{tu2022differentiable,liu2023multi,tu2024two,huo2023deep,huo2024deep,tu2025cross} and use 
pretrained CLIP~\cite{radford2021learning} image and text encoders, $E_I$ and $E_T$, to extract continuous features 
$\mathbf{x}_i=E_I(I_i)$ and $\mathbf{y}_i=E_T(T_i)$, respectively. CMH learns two mappings, $f_I$ and $f_T$, to transform 
these features into K-bit binary codes, $\mathbf{b}_i^I=f_I(\mathbf{x}_i)$ and $\mathbf{b}_i^T=f_T(\mathbf{y}_i)$, where 
$\mathbf{b}_i^I$,$\mathbf{b}_i^T\in\{-1,+1\}^{K}$. As shown in Fig.~\ref{Comparison_of_Framework_Diagrams}, existing methods 
typically learn continuous representations and apply binarization at the final stage. In contrast, SpikeBit generates hash 
bits through discrete spiking activity over time.

\noindent\textbf{Spiking neural networks.}
Spiking neural networks (SNNs)~\cite{cao2015spiking} transmit information through discrete spike events and use membrane 
potentials to model neuronal dynamics over time. We adopt the Parametric Leaky Integrate-and-Fire (PLIF) 
neuron~\cite{fang2021incorporating} as the basic computational unit $\mathcal{S}_{\mathrm{PLIF}}$, with its state updates defined as:
\begin{equation}
\begin{aligned}
H[t]&=U[t-1]+\frac{1}{\tau}\left(I[t]-(U[t-1]-U_{\mathrm{reset}})\right)\\
S[t]&=\Theta(H[t]-U_{\mathrm{th}})\\
U[t]&=H[t](1-S[t])+U_{\mathrm{reset}}S[t]
\end{aligned}
\end{equation}
where $H[t]$ is the membrane potential before spike firing, and $S[t]\in\{0,1\}$ records whether the neuron fires at time 
step $t$. $U_{\mathrm{th}}$ and $U_{\mathrm{reset}}$ denote the firing threshold and reset potential, respectively, and 
$\Theta(\cdot)$ is the Heaviside step function. The learnable parameter $\tau$ controls the rate at which the neuron 
integrates its input with the membrane state inherited from earlier time steps. The spike $S[t]$ represents the event 
emitted at the current step, whereas $U[t]$ carries accumulated information across time. 
\subsection{Theoretical motivation}\label{theomotiva}
We analyze how PLIF dynamics convert static continuous inputs into temporal spikes and how accumulated spike competition 
supports stable hash-bit decisions. 

\noindent\textbf{Proposition 1 (Temporal spike formation).} Consider a PLIF neuron with constant input $a$, initial 
potential $U[0]=U_{\mathrm{reset}}$, and hard reset. Let $\alpha=1-1/\tau\in(0,1)$ and $\theta=U_{\mathrm{th}}-U_{\mathrm{reset}}>0$. 
Under the firing rule $H[t]\geq U_{\mathrm{th}}$, the first-spike time $L(a)$ and spike count $C_{N_T}(a)$ satisfy:
\begin{equation} L(a)= \left\lceil \frac{\ln(1-\theta/a)}{\ln\alpha} \right\rceil \end{equation}
\begin{equation} C_{N_T}(a)= \begin{cases} \lfloor N_T/L(a)\rfloor, & a>\theta,\\ 0, & a\leq\theta \end{cases} \end{equation}
Here, $L(a)$ is non-increasing and $C_{N_T}(a)$ is non-decreasing with respect to $a$.

\noindent\textit{Proof.}
Before the first spike, the membrane potential follows $H[t]-U_{\mathrm{reset}} =\alpha(U[t-1]-U_{\mathrm{reset}})+a/\tau$. 
Since $U[0]=U_{\mathrm{reset}}$ and no reset occurs before the first firing threshold test, $H[t]-U_{\mathrm{reset}}=a(1-\alpha^t)$. 
Solving $a(1-\alpha^t)\geq\theta$ gives $L(a)$. Hard reset and constant input yield identical inter-spike intervals, giving $C_{N_T}(a)$. The monotonicity follows from the threshold condition.

\noindent\textcolor{red}{\underline{\textbf{Remark.}}}
PLIF dynamics transform constant continuous inputs into discrete temporal responses characterized by firing times and spike counts through membrane integration and reset.

\noindent\textbf{Proposition 2 (Stability of competitive hash-bit decisions).}
For modality $m$, let $Z_{k,+}^{m}[t],Z_{k,-}^{m}[t]\in\{0,1\}$ denote the positive and negative spike outputs for hash bit $k$. 
Define the accumulated difference $D_k^m=\sum_{t=1}^{N_T} (Z_{k,+}^{m}[t]-Z_{k,-}^{m}[t])$. The resulting bit is $b_k^m=+1$ if 
$D_k^m\geq0$ and $b_k^m=-1$ otherwise. If the per-step spike difference changes at most $r$ time steps, then:
\begin{equation}
|D_k^m|>2r
\quad\Longrightarrow\quad
\widetilde b_k^m=b_k^m
\end{equation}

\noindent\textit{Proof.}
Let $X_k^m[t]=Z_{k,+}^m[t]-Z_{k,-}^m[t]$ and let $\mathcal A$ denote the set of altered time steps, where $|\mathcal A|\leq r$. 
Since both $X_k^m[t]$ and $\widetilde X_k^m[t]$ belong to $\{-1,0,1\}$, the triangle inequality gives:
\begin{equation}
\begin{aligned}
|\widetilde D_k^m-D_k^m|
&\leq
\sum_{t\in\mathcal A}
|\widetilde X_k^m[t]-X_k^m[t]|\\
&\leq 2|\mathcal A|\leq 2r
\end{aligned}
\end{equation}
Thus, $|D_k^m|>2r$, the accumulated difference retains its sign, and the resulting hash bit remains unchanged.

\noindent\textcolor{red}{\underline{\textbf{Remark.}}}
A larger $|D_k^m|$ indicates a more stable hash-bit decision under spike perturbations. When both channels
remain silent, the bit defaults to $+1$ without spike support. Bit Activation Regularization is therefore
introduced to reduce such silent responses.

\noindent\textit{Detailed proofs are provided in the \textbf{supplementary} material.}
\begin{figure}[!ht]
    \centering
    \includegraphics[width=1.0\linewidth]{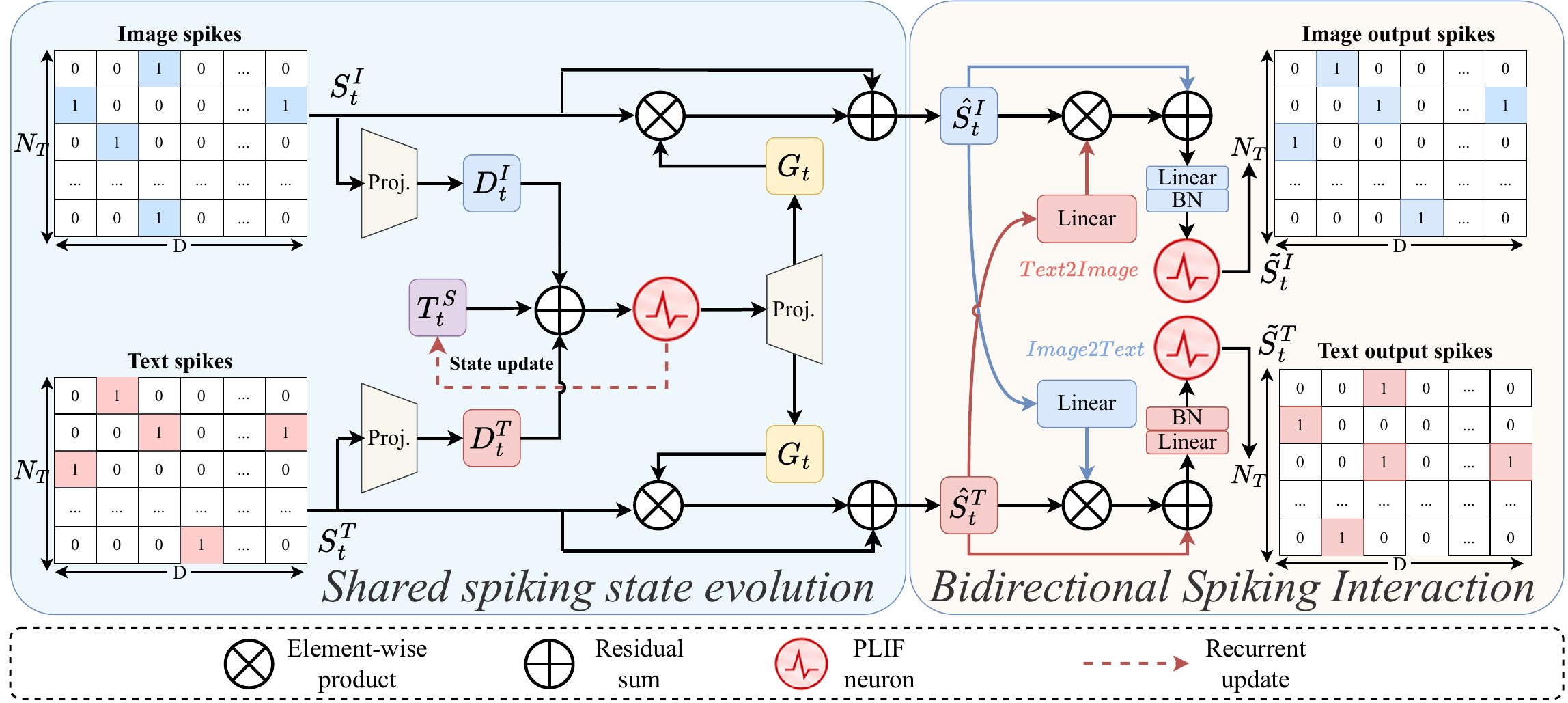} 
    \vspace{-18pt}
    \caption{
Overview of the cross-modal spiking interaction mechanism.
The shared spiking state evolves from image and text projections
$D_t^m=\phi_m(S_t^m)$ and the previous state
$T_t^s=\phi_R(h[t-1])$, where $m\in\{I,T\}$.
The resulting shared state generates modulation signals
$G_t=\phi_G(h[t])$ for both modalities.
Bidirectional interactions further exchange information
between image and text to produce the output representations.
}
    \label{fig2}
    \end{figure}
\subsection{SpikeBit Framework: Simple and efficient}
Without relying on complex self-attention, cross-attention, graph neural networks, or Mamba architectures, SpikeBit
enables efficient CMH through a simple spiking framework.

\noindent\textbf{Multi-step spike encoding.}
To obtain temporal spike representations for subsequent cross-modal interaction, we first repeat the image and text features over $N_T$
time steps~\cite{zheng2025spiliformer,lv2026toward}. The repeated features are then processed by separate linear projections, 
layer normalization, and PLIF neurons for each modality:
\begin{equation}
    \begin{aligned}
    S_I[t]&=\mathcal{S}_{\mathrm{PLIF}}\left(\mathrm{LN}_I(W_I \mathbf{x}[t]+b_I)\right)\\
    S_T[t]&=\mathcal{S}_{\mathrm{PLIF}}\left(\mathrm{LN}_T(W_T \mathbf{y}[t]+b_T)\right)
\end{aligned}
\end{equation}
where $W_m$ and $b_m$ denote the learnable projection matrix and bias for modality $m\in\{I,T\}$, respectively, and
$\mathrm{LN}_m$ denotes layer normalization. This process produces the spike sequences
$S_I,S_T\in\{0,1\}^{N_T\times D}$.

\noindent\textbf{Cross-modal spiking interaction.}
To establish semantic connections between image and text in the spiking domain, we introduce a cross-modal interaction mechanism
that combines shared state evolution with bidirectional spike modulation. As illustrated in Fig.~\ref{fig2}, this
mechanism first constructs a shared temporal state from both modalities and then enables bidirectional information exchange
to produce representations incorporating cross-modal information.

\noindent\textit{Shared spiking state evolution.}
Given the image and text spike sequences $S_I[t]$ and $S_T[t]$, we project them into a shared state space using separate mapping functions. The projected responses are combined with
the shared spike state from the previous time step to form the current input:
\begin{equation}
R[t]=\phi_F\left(
\phi_I(S_I[t])
+\phi_T(S_T[t])
+\phi_R(h[t-1])
\right)
\end{equation}
where $\phi_I$ and $\phi_T$ denote the image and text projection functions, respectively, $\phi_R$ incorporates the previous
shared state, and $\phi_F$ denotes the fusion mapping. The shared state is updated through the temporal dynamics of a PLIF neuron:
\begin{equation}
(U_h[t],h[t])
=
\mathcal{S}^{t}_{\mathrm{PLIF}}
\left(R[t],U_h[t-1]\right)
\end{equation}
where $U_h[t]$ and $h[t]$ represent the shared membrane potential and spike state, respectively.
This recurrent process integrates the current responses of both modalities with information from previous time steps.
\begin{figure}[!ht]
    \centering
    \includegraphics[width=0.8\linewidth]{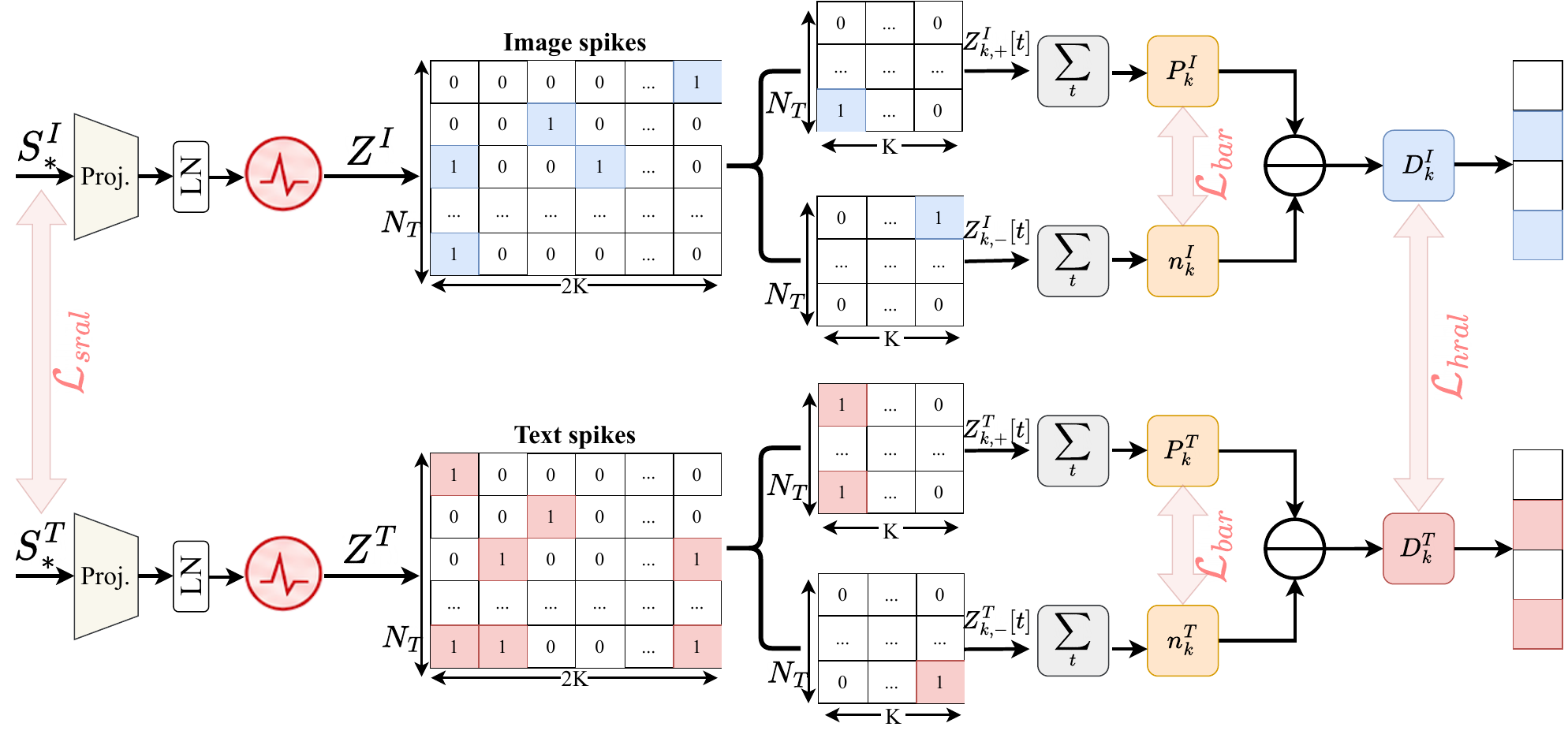} 
    \vspace{-10pt}
    \caption{Illustration of competitive hash-bit generation.
    Image and text representations are projected into $2K$
    channels and converted into spike sequences by PLIF neurons.
    Each positive-negative channel pair determines a hash bit
    through accumulated spike-count differences.}
    \label{fig3}
    \end{figure}

We feed the shared state back into both modalities through a channel modulation signal $G[t]=\phi_G(h[t])$:
\begin{equation}
\hat{S}_m[t]
=
\mathrm{BN}^{s}_m\left(
S_m[t]+S_m[t]\odot G[t]
\right)
\end{equation}
where $m\in\{I,T\}$ and $\odot$ denotes element-wise multiplication. This modulation allows both modalities to
adjust their channel responses according to the same temporal context while preserving their original spike information.

\noindent\textit{Bidirectional Spiking Interaction.}
Building on the shared state modulation, we further introduce
bidirectional information exchange, allowing each modality
to directly regulate the response of the other.
Let $\bar{m}$ denote the modality paired with $m$.
The interaction is formulated as:
\begin{equation}
\begin{aligned}
g_{\bar{m}\rightarrow m}[t]
&=W_{\bar{m}\rightarrow m}\hat{S}_{\bar{m}}[t]\\
R_{\bar{m}\rightarrow m}[t]
&=\hat{S}_m[t]
+\gamma\,\hat{S}_m[t]\odot g_{\bar{m}\rightarrow m}[t]\\
\tilde{S}_{\bar{m}\rightarrow m}[t]
&=\mathcal{S}_{\mathrm{PLIF}}\left(
\mathrm{BN}^{d}_m
\left(W_m^{d}R_{\bar{m}\rightarrow m}[t]\right)
\right)
\end{aligned}
\end{equation}

The source modality generates the modulation signal, while the target modality provides the response to be modulated.
Their element-wise interaction determines how the target channels are affected. The modulated responses are converted into interaction spikes by modality-specific PLIF neurons.

Before hash-bit generation, we estimate adaptive fusion gains for the shared and cross-modal interaction features from channel-wise relationships:
\begin{equation}
\begin{aligned}
&\Phi^m[t] = \operatorname{Stack}\big(
S_m[t], \hat{S}_m[t], \tilde{S}_{\bar{m}}[t],\\
&\qquad\qquad S_m[t]\odot\hat{S}_m[t],
S_m[t]\odot\tilde{S}_{\bar{m}}[t],\\
&\qquad\qquad |S_m[t]-\hat{S}_m[t]|,
|S_m[t]-\tilde{S}_{\bar{m}}[t]|
\big)\\
&[G_s^m[t],G_i^m[t]]
= 2\sigma\left(\Phi^m[t]W+b+E\right)
\end{aligned}
\label{eq:adaptive_spike_fusion}
\end{equation}

Here, $\operatorname{Stack}$ stacks the seven feature terms along a new trailing dimension, $W$ and $b$ denote the linear mapping parameters shared across channels, and $E$ is a learnable channel-specific bias. \(\sigma\) denotes the sigmoid function.

We combine the weighted features with the original response to form the joint representation $S_*^m[t]$:
\begin{equation}
    S_*^m[t] =
    \underbrace{S_m[t]}_{\text{Original}}
    +
    \underbrace{G_s^m[t]\odot\hat{S}_m[t]}_{\text{Shared}}
    +
    \underbrace{G_i^m[t]\odot\tilde{S}_{\bar{m}}[t]}_{\text{Interaction}}
    \label{eq:adaptive_fusion}
    \end{equation}

This representation drives the positive and negative hash channels, and the difference between their accumulated spike counts determines each hash bit at readout.

\noindent\textbf{Competitive hash-bit generation.}
To convert the temporal responses $S_*^m$ into bipolar codes for Hamming-space retrieval, we assign two readout channels
to each hash bit. A spike indicates only firing or silence. The paired channels instead take positive and negative roles
in the bit decision. As shown in Fig.~\ref{fig3}, a shared hash head projects the responses from each modality into
$2K$ channels. Layer normalization and PLIF neurons then produce the spike sequences:
\begin{equation}
Z^m=
\mathcal{S}_{\mathrm{PLIF}}
\left(\mathrm{LN}(W_h S_*^m)\right),
Z^m\in\{0,1\}^{N_T\times 2\times K}
\end{equation}

Each pair of channels corresponds to one hash bit. The positive and negative labels denote the readout roles of the channels, 
not the spike values. For the $k$-th bit, the difference between the accumulated
spike counts of the two channels defines the competition score and determines the bit value:
\begin{equation}
\begin{aligned}
D_k^m
&=\sum_{t=1}^{N_T}
\left(Z_{k,+}^m[t]-Z_{k,-}^m[t]\right)\\
b_k^m
&=
\begin{cases}
+1, & D_k^m\geq 0,\\
-1, & D_k^m<0
\end{cases}
\end{aligned}
\end{equation}

During training, the objective operates on the competition score vector $\mathbf{D}^m=[D_1^m,\ldots,D_K^m]$, while
the same vector determines the binary codes at inference. Thus, semantic learning directly shapes the firing responses
used for bit decisions without requiring a separate continuous hash representation independent of the readout process.

\noindent\textbf{Optimization objective.} When both channels of a hash bit remain silent throughout the time window, the competition score becomes zero. The tie-breaking rule then assigns the bit $+1$, although neither channel supports this decision. Let $p_{i,k}^m=\sum_t Z_{i,k,+}^m[t]$ and
$n_{i,k}^m=\sum_t Z_{i,k,-}^m[t]$ denote the accumulated spike counts of the positive and negative channels for the $k$-th
bit of sample $i$ in modality $m$, respectively. We introduce Bit Activation Regularization (BAR) to penalize
silent channel pairs:
\begin{equation} \mathcal{L}_{\mathrm{bar}} = \frac{1}{BK} \sum_{i=1}^{B} \sum_{k=1}^{K} \sum_{m\in\{I,T\}} 
\max\left( 0,1-p_{i,k}^m-n_{i,k}^m \right) \end{equation}

This regularization encourages at least one channel to fire without favoring either bit polarity.

We further introduce cross-modal contrastive learning before and after the hash readout.
Given paired image and text representations $X$ and $Y$ in a batch, we normalize and concatenate them into
$Z=[z_1,\ldots,z_{2B}] =[\mathrm{norm}(X);\mathrm{norm}(Y)]$. The bidirectional paired contrastive loss is defined as:
\begin{equation}
\mathcal{L}_{\mathrm{con}}(X,Y)
=
-\frac{1}{2B}
\sum_{i=1}^{2B}
\log
\frac{
\exp\left(\mathrm{sim}(z_i,z_i^+)/\tau_{\mathrm{con}}\right)
}{
\sum_{\substack{j\neq i}}^{2B}
\exp\left(\mathrm{sim}(z_i,z_j)/\tau_{\mathrm{con}}\right)
}
\end{equation}
where $z_i^+$ denotes the paired sample from the other modality, and $\mathrm{sim}(\cdot,\cdot)$ denotes cosine similarity.

Applying this loss to the competition scores $\mathbf{D}^I$ and $\mathbf{D}^T$ defines the Hash Readout Alignment Loss (HRAL). Applying it to the temporally averaged interaction responses
$\bar{\mathbf{s}}^m=\frac{1}{N_T}\sum_t S_*^m[t]$ defines the Spiking Representation Alignment Loss (SRAL). HRAL directly aligns the readout scores used for bit decisions,
while SRAL aligns the cross-modal representations before they enter the hash head.
The overall objective is:
\begin{equation}
\begin{aligned}
\mathcal{L}
=\underbrace{\mathcal{L}_{\mathrm{con}}(\mathbf{D}^I,\mathbf{D}^T)}
_{\mathcal{L}_{\mathrm{hral}}}
+\lambda_{\mathrm{sral}}
\underbrace{\mathcal{L}_{\mathrm{con}}(\bar{\mathbf{s}}^I,\bar{\mathbf{s}}^T)}
_{\mathcal{L}_{\mathrm{sral}}}
+\lambda_{\mathrm{bar}}\mathcal{L}_{\mathrm{bar}}
\end{aligned}
\end{equation}

During training, we use a surrogate gradient for the PLIF firing function in backpropagation.
\begin{table*}[!ht]
\centering
\caption{Cross-modal retrieval performance on MIRFLICKR, NUSWIDE,
and MSCOCO. The best unsupervised result is \textbf{bold} and the second-best is
\underline{underlined}.}
\vspace{-5pt}
\label{tab:main_retrieval}
\small
\begin{tabular}{@{}l*{4}{r}@{\hspace{9pt}}*{4}{r}@{\hspace{9pt}}*{4}{r}@{}}
\toprule
\multirow{2}{*}{\textbf{Method}}
& \multicolumn{4}{c}{\textbf{MIRFLICKR}}
& \multicolumn{4}{c}{\textbf{NUS-WIDE}}
& \multicolumn{4}{c}{\textbf{MS-COCO}} \\
\cmidrule(lr){2-5}\cmidrule(lr){6-9}\cmidrule(lr){10-13}
& 16 & 32 & 64 & 128
& 16 & 32 & 64 & 128
& 16 & 32 & 64 & 128 \\
\midrule
\multicolumn{13}{@{}l}{\retrref{Supervised reference}} \\
\retrref{DCMHT\cite{tu2022differentiable}}
& \retrref{91.67} & \retrref{91.17} & \retrref{92.14} & \retrref{90.52}
& \retrref{79.79} & \retrref{77.60} & \retrref{79.22} & \retrref{79.70}
& \retrref{74.50} & \retrref{76.73} & \retrref{76.45} & \retrref{73.51} \\
\retrref{DIMCH\cite{tu2025cross}}
& \retrref{92.85} & \retrref{93.04} & \retrref{94.07} & \retrref{93.01}
& \retrref{80.98} & \retrref{82.53} & \retrref{84.81} & \retrref{85.67}
& \retrref{85.72} & \retrref{88.42} & \retrref{89.61} & \retrref{88.82} \\
\retrref{DNPH\cite{huo2024deep}}
& \retrref{94.78} & \retrref{95.84} & \retrref{96.46} & \retrref{96.77}
& \retrref{82.76} & \retrref{84.30} & \retrref{84.07} & \retrref{83.06}
& \retrref{86.26} & \retrref{90.89} & \retrref{92.41} & \retrref{91.11} \\
\retrref{DSPH\cite{huo2023deep}}
& \retrref{95.41} & \retrref{95.84} & \retrref{96.46} & \retrref{96.77}
& \retrref{83.01} & \retrref{84.00} & \retrref{85.81} & \retrref{87.29}
& \retrref{88.54} & \retrref{92.51} & \retrref{94.37} & \retrref{95.44} \\
\retrref{TwDH\cite{tu2024two}}
& \retrref{94.86} & \retrref{95.37} & \retrref{96.41} & \retrref{--}
& \retrref{82.85} & \retrref{84.84} & \retrref{85.97} & \retrref{--}
& \retrref{90.29} & \retrref{93.81} & \retrref{94.99} & \retrref{--} \\
\retrref{Supervised mean}
& \retrref{93.91} & \retrref{94.25} & \retrref{95.11} & \retrref{94.27}
& \retrref{81.88} & \retrref{82.65} & \retrref{83.98} & \retrref{83.93}
& \retrref{85.06} & \retrref{88.47} & \retrref{89.57} & \retrref{87.22} \\

\multicolumn{13}{@{}l}{\textbf{Image-to-text retrieval (I$\to$T)}} \\
\addlinespace[2pt]
UCCH\cite{hu2022unsupervised}
& 88.55 & 91.43 & 92.44 & 93.16
& 77.85 & 80.82 & 82.25 & 83.55
& 78.07 & 87.62 & 88.35 & 92.57 \\
VTMUCH\cite{fan2025vision}
& \retrsecond{90.30} & 90.97 & 90.66 & 90.67
& 75.43 & 75.84 & 76.31 & 77.21
& 68.36 & 71.00 & 70.58 & 71.97 \\
BRAIN\cite{fu2026brainhash}
& 87.88 & 90.49 & 90.93 & 91.61
& 77.37 & 80.39 & 81.53 & 82.63
& 87.76 & 90.08 & 91.92 & 92.55 \\
UCMFH\cite{xia2023clip}
& 88.38 & 91.04 & 91.44 & 92.06
& 77.30 & 80.25 & 82.33 & 81.94
& 88.00 & 91.67 & 92.72 & 92.40 \\
DDSS\cite{huang2025dual}
& 89.64 & \retrsecond{92.20} & \retrsecond{93.12} & \retrsecond{93.46}
& \retrsecond{79.17} & \retrsecond{82.75} & \retrsecond{84.05} & \retrsecond{84.45}
& \retrsecond{90.27} & \retrsecond{93.04} & \retrsecond{93.73} & \retrbest{94.20} \\
\rowcolor{RetrievalOursBg}
\textbf{SpikeBit (Ours)}
& \retrbest{91.10} & \retrbest{92.97} & \retrbest{93.74} & \retrbest{94.37}
& \retrbest{80.35} & \retrbest{83.01} & \retrbest{84.25} & \retrbest{84.71}
& \retrbest{91.15} & \retrbest{93.15} & \retrbest{94.04} & \retrsecond{94.12} \\
\midrule
\multicolumn{13}{@{}l}{\retrref{Supervised reference}} \\
\retrref{DCMHT\cite{tu2022differentiable}}
& \retrref{87.37} & \retrref{88.66} & \retrref{90.51} & \retrref{90.34}
& \retrref{77.04} & \retrref{75.37} & \retrref{78.00} & \retrref{78.87}
& \retrref{73.46} & \retrref{77.47} & \retrref{75.19} & \retrref{73.79} \\
\retrref{DIMCH\cite{tu2025cross}}
& \retrref{91.08} & \retrref{92.10} & \retrref{92.49} & \retrref{91.51}
& \retrref{82.31} & \retrref{83.81} & \retrref{80.72} & \retrref{81.35}
& \retrref{86.18} & \retrref{88.76} & \retrref{89.45} & \retrref{89.45} \\
\retrref{DNPH\cite{huo2024deep}}
& \retrref{90.03} & \retrref{90.86} & \retrref{91.16} & \retrref{91.22}
& \retrref{80.14} & \retrref{82.03} & \retrref{82.86} & \retrref{83.06}
& \retrref{89.11} & \retrref{93.05} & \retrref{94.20} & \retrref{93.71} \\
\retrref{DSPH\cite{huo2023deep}}
& \retrref{89.82} & \retrref{90.86} & \retrref{91.16} & \retrref{91.22}
& \retrref{80.47} & \retrref{83.03} & \retrref{83.42} & \retrref{84.63}
& \retrref{89.89} & \retrref{93.70} & \retrref{95.38} & \retrref{96.64} \\
\retrref{TwDH\cite{tu2024two}}
& \retrref{90.00} & \retrref{90.82} & \retrref{90.90} & \retrref{--}
& \retrref{78.27} & \retrref{80.22} & \retrref{81.03} & \retrref{--}
& \retrref{92.52} & \retrref{95.10} & \retrref{96.13} & \retrref{--} \\
\retrref{Supervised mean}
& \retrref{89.66} & \retrref{90.66} & \retrref{91.24} & \retrref{91.07}
& \retrref{79.65} & \retrref{80.89} & \retrref{81.21} & \retrref{81.98}
& \retrref{86.23} & \retrref{89.62} & \retrref{90.07} & \retrref{88.40} \\
\multicolumn{13}{@{}l}{\textbf{Text-to-image retrieval (T$\to$I)}} \\
\addlinespace[2pt]
UCCH\cite{hu2022unsupervised}
& 85.97 & 88.06 & 89.03 & 89.71
& 75.14 & 78.43 & 79.59 & 80.75
& 77.91 & 87.59 & 88.48 & 92.43 \\
VTMUCH\cite{fan2025vision}
& 86.45 & 86.38 & 86.90 & 87.50
& 74.19 & 74.78 & 74.61 & 75.01
& 69.35 & 69.65 & 71.37 & 72.21 \\
BRAIN\cite{fu2026brainhash}
& 85.95 & 87.94 & 88.27 & 88.89
& 75.96 & 79.02 & 80.13 & 80.93
& 87.96 & 90.88 & 92.47 & 92.96 \\
UCMFH\cite{xia2023clip}
& 87.67 & 90.38 & 91.26 & 91.21
& 75.65 & 79.51 & 81.09 & 81.28
& 88.23 & 91.83 & 92.66 & 92.42 \\
DDSS\cite{huang2025dual}
& \retrsecond{89.43} & \retrsecond{92.28} & \retrsecond{93.09} & \retrsecond{93.47}
& \retrsecond{78.71} & \retrsecond{82.87} & \retrbest{84.24} & \retrsecond{84.29}
& \retrsecond{90.45} & \retrsecond{93.00} & \retrsecond{93.77} & \retrbest{94.30} \\
\rowcolor{RetrievalOursBg}
\textbf{SpikeBit (Ours)}
& \retrbest{90.98} & \retrbest{92.96} & \retrbest{93.84} & \retrbest{94.50}
& \retrbest{80.37} & \retrbest{83.10} & \retrsecond{84.22} & \retrbest{84.65}
& \retrbest{91.18} & \retrbest{93.16} & \retrbest{94.09} & \retrsecond{94.11} \\
\bottomrule
\end{tabular}%
\end{table*}
\begin{table*}[!ht]
    \centering
    \caption{Cross-dataset generalization on MIRFLICKR, NUSWIDE, and MSCOCO. The best and second-best results are \textbf{bold} and \underline{underlined}, respectively.}
\vspace{-5pt}
    \label{tab:cross_dataset_generalization}
    \small
    \setlength{\tabcolsep}{3.5pt}
    \begin{tabular}{@{}clc*{11}{c}@{}}
    \toprule
    & \multirow{2}{*}{\textbf{Method}}
    & \multicolumn{4}{c}{\textbf{MIRFLICKR $\to$ MS-COCO}}
    & \multicolumn{4}{c}{\textbf{NUS-WIDE $\to$ MS-COCO}}
    & \multicolumn{4}{c}{\textbf{MIRFLICKR $\to$ NUS-WIDE}} \\
    \cmidrule(lr){3-6}\cmidrule(lr){7-10}\cmidrule(lr){11-14}
    & & 16 & 32 & 64 & 128
        & 16 & 32 & 64 & 128
        & 16 & 32 & 64 & 128 \\
    \midrule
    
    \multirow{6}{*}{\transferdir{I$\to$T}}
    & UCCH\cite{hu2022unsupervised}
    & 62.34 & 68.14 & 70.47 & 75.26
    & 64.20 & 70.50 & 73.65 & 77.25
    & 72.19 & 76.03 & 78.37 & 80.26 \\
    & VTMUCH\cite{fan2025vision}
    & 59.24 & 62.31 & 61.91 & 60.03
    & 55.56 & 53.85 & 58.41 & 59.53
    & 72.18 & 72.60 & 73.27 & 71.86 \\
    & BRAIN\cite{fu2026brainhash}
    & 57.82 & 66.91 & 65.93 & 68.72
    & 65.71 & 71.34 & 73.61 & 75.67
    & 70.83 & 72.94 & 74.77 & 75.58 \\
    & UCMFH\cite{xia2023clip}
    & 56.78 & 71.42 & 72.67 & 77.65
    & 55.93 & 67.55 & 73.12 & 74.75
    & 70.44 & 77.08 & 77.29 & 78.72 \\
    & DDSS\cite{huang2025dual}
    & \transfersecond{66.28} & \transfersecond{72.46}
    & \transfersecond{77.23} & \transfersecond{78.40}
    & \transfersecond{67.85} & \transfersecond{76.48}
    & \transfersecond{80.98} & \transfersecond{82.75}
    & \transfersecond{75.64} & \transfersecond{79.00}
    & \transfersecond{80.80} & \transfersecond{81.41} \\
    & \cellcolor{TransferOursBg}\textbf{SpikeBit (Ours)}
    & \cellcolor{TransferOursBg}\transferbest{68.26}
    & \cellcolor{TransferOursBg}\transferbest{75.98}
    & \cellcolor{TransferOursBg}\transferbest{80.19}
    & \cellcolor{TransferOursBg}\transferbest{82.81}
    & \cellcolor{TransferOursBg}\transferbest{71.44}
    & \cellcolor{TransferOursBg}\transferbest{78.71}
    & \cellcolor{TransferOursBg}\transferbest{81.98}
    & \cellcolor{TransferOursBg}\transferbest{83.17}
    & \cellcolor{TransferOursBg}\transferbest{76.31}
    & \cellcolor{TransferOursBg}\transferbest{80.19}
    & \cellcolor{TransferOursBg}\transferbest{82.85}
    & \cellcolor{TransferOursBg}\transferbest{83.72} \\
    \midrule
    
    \multirow{6}{*}{\transferdir{T$\to$I}}
    & UCCH\cite{hu2022unsupervised}
    & 64.48 & 69.89 & 70.62 & 75.62
    & 63.26 & 69.32 & 74.31 & 77.96
    & 69.77 & 70.35 & 76.14 & 77.65 \\
    & VTMUCH\cite{fan2025vision}
    & 60.01 & 61.16 & 61.92 & 62.30
    & 62.25 & 62.90 & 63.08 & 62.62
    & 68.06 & 71.24 & 72.29 & 70.94 \\
    & BRAIN\cite{fu2026brainhash}
    & 59.35 & 63.88 & 68.46 & 66.95
    & 63.72 & 70.97 & 72.83 & 74.24
    & 69.05 & 71.54 & 74.99 & 75.53 \\
    & UCMFH\cite{xia2023clip}
    & 55.38 & 69.65 & 72.80 & 77.85
    & 57.37 & 67.39 & 73.99 & 74.00
    & 67.43 & 75.81 & 77.44 & 78.34 \\
    & DDSS\cite{huang2025dual}
    & \transfersecond{66.91} & \transfersecond{72.41}
    & \transfersecond{77.69} & \transfersecond{79.14}
    & \transfersecond{69.08} & \transfersecond{76.37}
    & \transfersecond{80.92} & \transfersecond{82.51}
    & \transfersecond{75.22} & \transfersecond{78.98}
    & \transfersecond{81.50} & \transfersecond{81.91} \\
    & \cellcolor{TransferOursBg}\textbf{SpikeBit (Ours)}
    & \cellcolor{TransferOursBg}\transferbest{68.76}
    & \cellcolor{TransferOursBg}\transferbest{75.90}
    & \cellcolor{TransferOursBg}\transferbest{80.30}
    & \cellcolor{TransferOursBg}\transferbest{82.97}
    & \cellcolor{TransferOursBg}\transferbest{70.76}
    & \cellcolor{TransferOursBg}\transferbest{79.00}
    & \cellcolor{TransferOursBg}\transferbest{81.83}
    & \cellcolor{TransferOursBg}\transferbest{83.13}
    & \cellcolor{TransferOursBg}\transferbest{75.74}
    & \cellcolor{TransferOursBg}\transferbest{80.29}
    & \cellcolor{TransferOursBg}\transferbest{82.94}
    & \cellcolor{TransferOursBg}\transferbest{83.60} \\
    
    \midrule
    \addlinespace[2pt]
    & \multirow{2}{*}{\textbf{Method}}
    & \multicolumn{4}{c}{\textbf{MS-COCO $\to$ MIRFLICKR}}
    & \multicolumn{4}{c}{\textbf{MS-COCO $\to$ NUS-WIDE}}
    & \multicolumn{4}{c}{\textbf{NUS-WIDE $\to$ MIRFLICKR}} \\
    \cmidrule(lr){3-6}\cmidrule(lr){7-10}\cmidrule(lr){11-14}
    & & 16 & 32 & 64 & 128
        & 16 & 32 & 64 & 128
        & 16 & 32 & 64 & 128 \\
    \midrule
    
    \multirow{6}{*}{\transferdir{I$\to$T}}
    & UCCH\cite{hu2022unsupervised}
    & 85.41 & 88.77 & 90.34 & \transfersecond{92.03}
    & 69.32 & 75.06 & 76.89 & 76.39
    & 86.09 & 88.39 & 89.19 & 91.06 \\
    & VTMUCH\cite{fan2025vision}
    & 86.76 & 87.52 & 88.73 & 87.77
    & 69.78 & 71.38 & 70.87 & 70.72
    & 82.72 & 83.07 & 84.63 & 85.96 \\
    & BRAIN\cite{fu2026brainhash}
    & 86.84 & 89.05 & 90.63 & 91.64
    & 73.58 & 76.98 & 78.95 & 79.41
    & 84.80 & 87.20 & 88.91 & 89.49 \\
    & UCMFH\cite{xia2023clip}
    & 79.96 & 84.56 & 85.74 & 89.39
    & 66.13 & 75.57 & 77.61 & 79.92
    & 75.76 & 82.44 & 85.68 & 86.78 \\
    & DDSS\cite{huang2025dual}
    & \transfersecond{87.99} & \transfersecond{91.24}
    & \transferbest{92.93} & \transferbest{93.84}
    & \transfersecond{75.92} & \transferbest{81.80}
    & \transferbest{83.52} & \transferbest{84.10}
    & \transfersecond{86.90} & \transferbest{89.61}
    & \transfersecond{91.04} & \transfersecond{91.83} \\
    & \cellcolor{TransferOursBg}\textbf{SpikeBit (Ours)}
    & \cellcolor{TransferOursBg}\transferbest{88.65}
    & \cellcolor{TransferOursBg}\transferbest{91.70}
    & \cellcolor{TransferOursBg}\transfersecond{90.90}
    & \cellcolor{TransferOursBg}91.98
    & \cellcolor{TransferOursBg}\transferbest{76.39}
    & \cellcolor{TransferOursBg}\transfersecond{80.08}
    & \cellcolor{TransferOursBg}\transfersecond{79.64}
    & \cellcolor{TransferOursBg}\transfersecond{80.87}
    & \cellcolor{TransferOursBg}\transferbest{87.19}
    & \cellcolor{TransferOursBg}\transfersecond{89.08}
    & \cellcolor{TransferOursBg}\transferbest{91.83}
    & \cellcolor{TransferOursBg}\transferbest{92.47} \\
    \midrule
    
    \multirow{6}{*}{\transferdir{T$\to$I}}
    & UCCH\cite{hu2022unsupervised}
    & 75.58 & 77.70 & 77.78 & 80.53
    & 57.76 & 62.61 & 64.49 & 68.35
    & 81.42 & 85.42 & 86.91 & 87.45 \\
    & VTMUCH\cite{fan2025vision}
    & 83.58 & 78.87 & 79.88 & 81.00
    & 68.84 & 66.82 & 66.11 & 68.13
    & 83.38 & 82.44 & 82.81 & 83.41 \\
    & BRAIN\cite{fu2026brainhash}
    & 79.34 & 81.79 & 79.91 & \transferbest{92.80}
    & 65.91 & 69.32 & 70.56 & 72.90
    & 83.02 & 85.87 & 86.99 & 86.79 \\
    & UCMFH\cite{xia2023clip}
    & 78.74 & 83.59 & 84.60 & 88.43
    & 66.30 & 75.23 & 75.31 & 78.79
    & 74.12 & 81.21 & 84.43 & 85.67 \\
    & DDSS\cite{huang2025dual}
    & \transfersecond{85.61} & \transfersecond{89.41}
    & \transfersecond{90.09} & 91.95
    & \transfersecond{73.16} & \transferbest{80.02}
    & \transferbest{80.88} & \transferbest{82.95}
    & \transfersecond{86.59} & \transfersecond{89.08}
    & \transfersecond{90.69} & \transfersecond{91.22} \\
    & \cellcolor{TransferOursBg}\textbf{SpikeBit (Ours)}
    & \cellcolor{TransferOursBg}\transferbest{87.95}
    & \cellcolor{TransferOursBg}\transferbest{91.89}
    & \cellcolor{TransferOursBg}\transferbest{90.93}
    & \cellcolor{TransferOursBg}\transfersecond{91.99}
    & \cellcolor{TransferOursBg}\transferbest{75.49}
    & \cellcolor{TransferOursBg}\transfersecond{79.79}
    & \cellcolor{TransferOursBg}\transfersecond{79.63}
    & \cellcolor{TransferOursBg}\transfersecond{81.09}
    & \cellcolor{TransferOursBg}\transferbest{87.23}
    & \cellcolor{TransferOursBg}\transferbest{89.13}
    & \cellcolor{TransferOursBg}\transferbest{91.82}
    & \cellcolor{TransferOursBg}\transferbest{92.49} \\
    \bottomrule
    \end{tabular}
    \end{table*}
\section{Experiments}
We evaluate SpikeBit by addressing four questions:
\begin{itemize}
    \item \textbf{$Q_1$ Retrieval accuracy (\S~\ref{retriacc}).}
    How does SpikeBit compare with existing methods in bidirectional retrieval across different code lengths?
    \item \textbf{$Q_2$ Cross-domain generalization (\S~\ref{crossdom}).}
    How well does SpikeBit generalize across datasets?
    \item \textbf{$Q_3$ Noise robustness (\S~\ref{noiserobu}).}
    How does feature noise affect retrieval accuracy?
    \item \textbf{$Q_4$ Ablation and overhead (\S~\ref{ablover}).}
    How do the key design choices contribute to performance, and what is the computational cost of hash-bit generation?
\end{itemize}
\subsection{Experimental settings}
\textbf{Datasets.} 
We use MIRFLICKR~\cite{huiskes2008mir}, NUSWIDE~\cite{chua2009nus}, and MSCOCO~\cite{lin2014microsoft} as the main evaluation benchmarks, containing 24,581, 195,834, and 122,218 image-text pairs, respectively. For each dataset, we randomly select 5,000 pairs as the query set and use the remaining pairs as the retrieval set, from which we randomly select 10,000 pairs as the training set.

\noindent\textbf{Baselines.}
We compare SpikeBit with five unsupervised methods: UCCH~\cite{hu2022unsupervised}, VTMUCH~\cite{fan2025vision}, BRAIN\cite{fu2026brainhash}, UCMFH~\cite{xia2023clip}, and DDSS~\cite{huang2025dual}. We reimplement all baselines using the same data scale and feature extraction network. Among them, BRAIN also employs a spiking mechanism. To evaluate the performance gap with supervised methods, we also reimplement DCMHT~\cite{tu2022differentiable}, DIMCH~\cite{tu2025cross}, DNPH~\cite{huo2024deep}, DSPH~\cite{huo2023deep}, and TwDH~\cite{tu2024two}.

\noindent\textbf{Evaluation metrics.} Mean average precision (mAP) serves as the primary evaluation metric, while Top-N precision, parameters, operations, estimated energy, and encoding time are used to comprehensively evaluate model performance.

\noindent\textbf{Implementation details.} All compared methods use the visual encoder of CLIP ViT-B/32 and its corresponding text encoder to extract image and text features, respectively. The batch size is set to 128. The model is trained using the Adam optimizer with a learning rate of $5\times10^{-3}$.  
SpikeBit is evaluated under the paired-input protocol adopted by UCMFH~\cite{xia2023clip}, DDSS~\cite{huang2025dual}, DAJSH~\cite{xu2025dual}, SCTH~\cite{yang2026stationary}, and USFTH~\cite{yang2025unsupervised}. All experiments for SpikeBit and the reproduced baselines are conducted on a single NVIDIA RTX 4090 GPU.
\subsection{Retrieval accuracy}\label{retriacc}
Table~\ref{tab:main_retrieval} shows that SpikeBit achieves the best unsupervised results in 21 of 24 comparisons and leads across all code lengths and retrieval directions on MIRFLICKR. Compared with BRAIN, which also employs a spiking mechanism, the advantage of SpikeBit suggests the potential impact of different hash readout strategies. BRAIN binarizes continuous membrane potentials using a sign function, retaining the separation between continuous representation learning and binary decisions. In contrast, SpikeBit determines bit polarity based on the difference in accumulated spike counts between positive and negative channels. By directly aligning this decision variable during training, SpikeBit reduces the risk of Hamming bit mismatches caused by continuous representations falling on opposite sides of a threshold. Furthermore, SpikeBit exceeds the supervised mean in 20 comparisons under the same settings but still falls short of the best supervised results for image-to-text retrieval on MIRFLICKR and text-to-image retrieval on MSCOCO.
\subsection{Cross-domain generalization}\label{crossdom}
Table~\ref{tab:cross_dataset_generalization} shows that SpikeBit achieves the best performance in most cross-domain settings. UCMFH and DDSS enhance continuous representations through attention or neighborhood structure modeling, whereas SpikeBit incorporates input-dependent modulation directly into spike generation and competitive readout. A potential advantage is that intermediate firing trajectories can vary with target-domain inputs, allowing some domain-related response changes to be accommodated without inducing bit flips. Consistent with Proposition 2 in §\ref{theomotiva}, this tolerance to variations in intermediate responses offers a way to combine input adaptability with code stability.
\subsection{Noise robustness}\label{noiserobu}
\begin{figure}[!ht]
    \centering
    \includegraphics[width=0.9\linewidth]{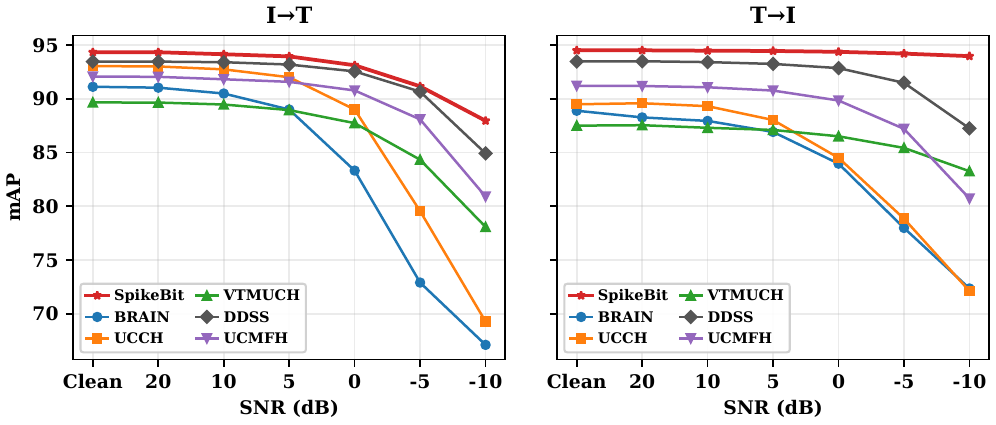}
    \vspace{-10pt}
    \caption{Robustness analysis under noise perturbations. }
    \label{fig:noise_robustness_mirflickr}
    \end{figure}

SpikeBit maintains the best performance across all noise levels, and Proposition 2 in §\ref{theomotiva} offers a possible explanation for this observation in Fig.~\ref{fig:noise_robustness_mirflickr}. When $|D_k^m|>2r$, temporal competition may preserve some bit decisions, thereby limiting the impact of noise on Hamming-distance rankings. Local changes in spiking activity therefore do not necessarily alter the final decisions, which may contribute to the robustness of competitive hash readout.
\begin{table}[!ht]
    \centering
    \caption{Module ablation on MIRFLICKR.
    CHBG denotes Competitive Hash-Bit Generation, SSSE denotes Shared Spiking
    State Evolution, and BSI denotes Bidirectional Spiking Interaction.}
\vspace{-5pt}
    \label{tab:module_ablation}
    \small
    \setlength{\tabcolsep}{2.5pt}
    \begin{tabular}{@{}cccccccc@{}}
        \toprule
        \multicolumn{4}{c}{\textbf{Modules}}
        & \multicolumn{2}{c}{I$\to$T}
        & \multicolumn{2}{c}{T$\to$I} \\
        \cmidrule(lr){1-4}\cmidrule(lr){5-6}\cmidrule(lr){7-8}
        Baseline & CHBG & SSSE & BSI & 64 & 128 & 64 & 128 \\
    \midrule
    \checkmark &            &            &
    & 90.08 & 91.69 & 87.26 & 88.49 \\
    \checkmark & \checkmark &            &
    & \underline{93.61} & \underline{93.94} & \underline{92.78} & \underline{93.42} \\
    \checkmark & \checkmark & \checkmark &
    & 91.17 & 91.12 & 91.19 & 90.74 \\
    \checkmark & \checkmark &            & \checkmark
    & 90.09 & 91.58 & 89.98 & 91.54 \\
    \midrule
    \checkmark & \checkmark & \checkmark & \checkmark
    & \textbf{93.74} & \textbf{94.37} & \textbf{93.84} & \textbf{94.50} \\
    \bottomrule
    \end{tabular}
    \end{table}
\subsection{Ablation and overhead}\label{ablover}
\textbf{Module effectiveness.} The baseline uses a conventional hash head for a fair comparison. Table~\ref{tab:module_ablation} shows that the gains from CHBG support generating hash bits directly from temporal responses through spike competition. Adding SSSE or BSI alone does not consistently improve performance. Combining them gives the best results, suggesting that the two modules depend on each other. 

This may be because SSSE provides a temporal context shared across modalities. BSI uses this context for bidirectional modulation and coordinates information exchange with shared state evolution. Removing either module may weaken this joint effect and affect hash-bit generation.

\begin{figure}[!ht]
    \centering
    \includegraphics[width=1\linewidth]{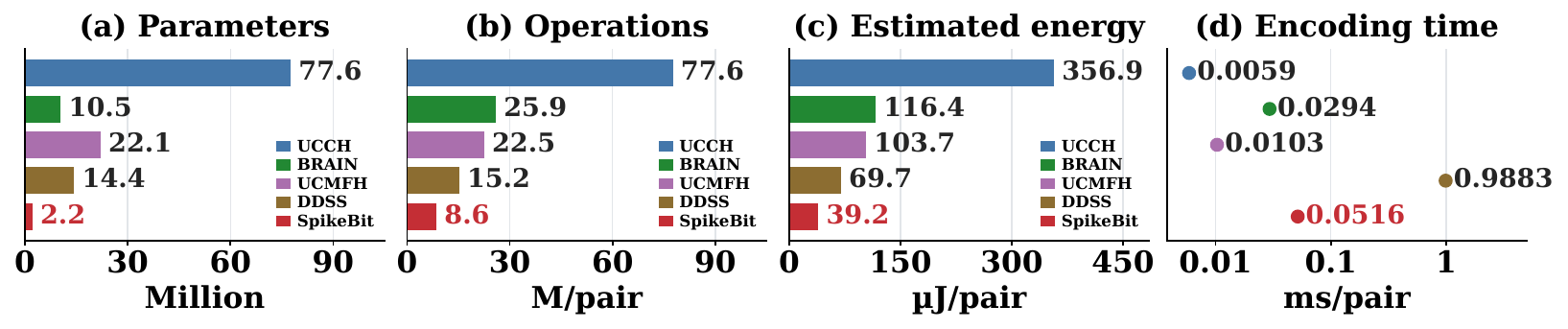} 
    \vspace{-20pt}
    \caption{Computational efficiency comparison.}
    \label{fig:efficiency_1x4}
    \end{figure}
\noindent\textbf{Overhead.} As shown in Fig.~\ref{fig:efficiency_1x4}, SpikeBit reduces parameter count, operation count, and estimated energy~\cite{lv2025spikebert, zhou2022spikformer} consumption by 79.1\%, 43.4\%, and 43.8\%, respectively, relative to the lowest baseline value for each metric. Dense ANN operators benefit from optimized GPU matrix computation kernels, whereas temporal dependencies in spiking states limit parallel execution across time steps~\cite{fang2023spikingjelly}. Fewer operations therefore do not necessarily translate into shorter encoding times. Nevertheless, SpikeBit requires only 0.0516 ms to encode each sample pair, approximately $\frac{1}{19}$ of the time required by DDSS, maintaining low encoding latency while substantially reducing resource requirements.
\begin{table}[!ht]
    \centering
    \caption{Effectiveness of each loss term.}
\vspace{-5pt}
    \label{tab:loss_ablation}
    \small
    \begin{tabular}{@{}ccccccc@{}}
    \toprule
    \multicolumn{3}{c}{\textbf{Loss terms}}
    & \multicolumn{2}{c}{I$\to$T}
    & \multicolumn{2}{c}{T$\to$I} \\
    \cmidrule(lr){1-3}\cmidrule(lr){4-5}\cmidrule(lr){6-7}
    $\mathcal{L}_{\mathrm{hral}}$
    & $\mathcal{L}_{\mathrm{sral}}$
    & $\mathcal{L}_{\mathrm{bar}}$
    & 64 & 128 & 64 & 128 \\
    \midrule
    \checkmark &            &
    & 92.73 & 93.89 & 92.68 & 93.85 \\
    \checkmark & \checkmark &
    & 92.74 & \underline{94.13}
    & 92.76 & \underline{94.10} \\
    \checkmark &            & \checkmark
    & \underline{93.64} & 93.92 & \underline{93.64} & 93.91 \\
    \midrule
    \checkmark & \checkmark & \checkmark
    & \textbf{93.74} & \textbf{94.37} & \textbf{93.84} & \textbf{94.50} \\
    \bottomrule
    \end{tabular}
    \end{table}
\noindent\textbf{Loss effectiveness.} Table~\ref{tab:loss_ablation} shows that combining all three losses yields the best performance, supporting their complementary roles. HRAL directly aligns the accumulated spike count differences that determine bit polarity, but the alignment of the overall readout vectors may not sufficiently constrain intermediate representations or firing activity for individual bits. SRAL strengthens cross-modal semantic consistency before readout, while BAR penalizes simultaneous silence in both channels, reducing default positive-bit assignments when no spikes occur. Together, they complement HRAL through representation alignment and bit activation, improving hash-bit decisions.

\begin{table}[!ht]
    \centering
    \caption{Ablations of hash-head design and spiking units. \textcircled{3} and \textcircled{5} represent the complete SpikeBit model. Group 1 varies only the hash head, while Group 2 replaces all spiking neurons with ANN units.}
\vspace{-5pt}
    \label{tab:head_and_spiking_ablation}
    \small
    \begin{tabular}{@{}clcccc@{}}
    \toprule
    & \multirow{2}{*}{\textbf{Configuration}}
    & \multicolumn{2}{c}{I$\to$T}
    & \multicolumn{2}{c}{T$\to$I} \\
    \cmidrule(lr){3-4}\cmidrule(lr){5-6}
    & & 64 & 128 & 64 & 128 \\
    \midrule
    \multirow{4}{*}{\rotatebox[origin=c]{90}{\textbf{Group 1}}}
& \multicolumn{5}{l}{\textit{Hash-head design}} \\
& \textcircled{1}Conventional head & 93.00 & 93.16 & 93.05 & 93.20 \\
& \textcircled{2}Simple spiking head & 91.18 & 91.32 & 91.15 & 91.02 \\
& \textcircled{3}CHBG spiking head
& \textbf{93.74} & \textbf{94.37} & \textbf{93.84} & \textbf{94.50} \\
\midrule
\multirow{3}{*}{\rotatebox[origin=c]{90}{\textbf{Group 2}}}
& \multicolumn{5}{l}{\textit{Spiking-unit ablation}} \\
& \textcircled{4}ANN counterpart & 91.91 & 92.74 & 92.05 & 92.72 \\
& \textcircled{5}Full spiking model
& \textbf{93.74} & \textbf{94.37} & \textbf{93.84} & \textbf{94.50} \\
\bottomrule
    \end{tabular}
    \end{table}
\noindent\textbf{Mechanism effectiveness.}  
Table~\ref{tab:head_and_spiking_ablation} shows that the simple spiking head (\textcircled{2}) underperforms the conventional head (\textcircled{1}), while the competitive head (\textcircled{3}) performs best. This suggests that organizing discrete firing into competition between bit polarities allows temporal responses and semantic optimization to jointly influence hash-bit decisions. In Group 2, we replace all PLIF neurons with ANN units while preserving the overall architecture, dual-channel competitive readout, and optimization objectives. The ANN counterpart (\textcircled{4}) consistently underperforms the full spiking model (\textcircled{5}), indicating that the performance gains cannot be attributed solely to the dual-channel competition mechanism. State evolution and cross-modal interactions regulate firing responses, while competitive readout converts these responses into hash bits, integrating representation learning and discrete decisions within a unified spiking process. 
\section{Conclusion}
This work investigates two questions in cross-modal hashing: how to determine hash bits and how image-text semantics can jointly guide bit generation. We determine bits through temporal competition between positive and negative spikes and propose SpikeBit, a lightweight unsupervised framework where cross-modal interactions regulate the competing firing responses. This design transforms spikes from information carriers into the basis for hash-bit generation. Experiments show that SpikeBit achieves leading retrieval performance with fewer resources, strong cross-domain generalization, and robustness to noise. 
\section*{LLM usage statement}
OpenAI's ChatGPT was used to assist with language translation, grammatical corrections, and improving the clarity of the manuscript. All AI-assisted revisions were reviewed and verified by the authors.
\section*{Acknowledgments}
This work is supported by the National Natural Science Foundation of China (62376106) and the Science and Technology Development Plan of Jilin Province (20250102212JC).

{
    \small
    \bibliographystyle{ieeenat_fullname}
    \bibliography{main}
}

\clearpage
\setcounter{page}{1}
\maketitlesupplementary

\setcounter{section}{0}
\renewcommand{\thesection}{S\arabic{section}}

\setcounter{equation}{0}
\renewcommand{\theequation}{S\arabic{equation}}

\setcounter{figure}{0}
\renewcommand{\thefigure}{S\arabic{figure}}

\setcounter{table}{0}
\renewcommand{\thetable}{S\arabic{table}}

\section{Overview}
We organize the supplementary material into five parts:
\begin{itemize}
\item \textbf{Theoretical proofs.}(\cref{theoproof}) We provide detailed proofs of the two propositions in the Theoretical motivation section (\cref{theomotiva}) of the main paper.
\item \textbf{Top-$N$ precision comparison.}(\cref{topn}) We compare SpikeBit with baseline methods in terms of bidirectional retrieval precision across different numbers of returned results.
\item \textbf{Time window length.}(\cref{timewind}) We analyze the effect of time window length on retrieval performance.
\item \textbf{Loss weights.}(\cref{lossweigh}) We compare retrieval performance across different weight combinations to determine the final loss weights.
\item \textbf{Qualitative comparison.}(\cref{quacompa}) We present retrieval results from SpikeBit and baseline methods to examine semantic matching and failure cases.
\end{itemize}
\section{Theoretical proofs}\label{theoproof}
This section provides detailed proofs of the two propositions in \cref{theomotiva} of the main paper. $N_T\in\mathbb N_+$ denotes the time window length. The neuronal time constant ($\tau$), firing threshold ($U_{\mathrm{th}}$), and reset potential ($U_{\mathrm{reset}}$) remain fixed throughout the time sequence under consideration. $r$ counts the number of time steps at which the spike difference changes.
\subsection{Proof 1: Temporal spike formation}
Consider a PLIF neuron with constant input $I[t]=a\in\mathbb R$ and initial state $U[0]=U_{\mathrm{reset}}$, using a hard reset and firing whenever $H[t]\geq U_{\mathrm{th}}$. Let:
\begin{equation}
\alpha=1-\frac{1}{\tau}\in(0,1),
\theta=U_{\mathrm{th}}-U_{\mathrm{reset}}>0
\end{equation}

Define the first-spike time as:
\begin{equation}
L(a)=\inf\left\{t\in\mathbb N_+:S[t]=1\right\},
\inf\varnothing=+\infty
\end{equation}

Then:
\begin{equation}
L(a)=
\begin{cases}
\displaystyle
\left\lceil
\frac{\ln(1-\theta/a)}{\ln\alpha}
\right\rceil,
& a>\theta\\[2mm]
+\infty,
& a\leq\theta
\end{cases}
\end{equation}

The spike count within the time window satisfies:
\begin{equation}
C_{N_T}(a)
=
\sum_{t=1}^{N_T}S[t]
=
\begin{cases}
\displaystyle
\left\lfloor\frac{N_T}{L(a)}\right\rfloor,
& a>\theta\\[2mm]
0,
& a\leq\theta
\end{cases}
\end{equation}

With all other parameters fixed, $L(a)$ is non-increasing in $a$, and $C_{N_T}(a)$ is non-decreasing in $a$.

\noindent\textbf{Proof.}
We first determine the first-spike time and the spike count, and then establish monotonicity.

\noindent\textbf{(1) First-spike time.}

The PLIF update rule gives:
\begin{equation}
    H[t]-U_{\mathrm{reset}}
    =
    \underbrace{
    \alpha\bigl(U[t-1]-U_{\mathrm{reset}}\bigr)
    }_{\text{Leaky membrane potential}}
    +
    \underbrace{
    \frac{a}{\tau}
    }_{\text{Current input}}
    \end{equation}

Define an auxiliary sequence without reset:
\begin{equation}
v[0]=0,
v[t]=\alpha v[t-1]+\frac{a}{\tau},
t\geq1
\end{equation}

Expanding the recurrence yields:
\begin{equation}
\begin{aligned}
v[t]
&=\alpha^t v[0]+\frac{a}{\tau}\sum_{j=0}^{t-1}\alpha^j\\
&=\frac{a}{\tau}\frac{1-\alpha^t}{1-\alpha}\\
&=a(1-\alpha^t)
\end{aligned}
\end{equation}

As long as no spike occurs at any preceding time step, the neuron undergoes no reset, and its membrane potential before threshold checking satisfies:
\begin{equation}
H[t]-U_{\mathrm{reset}}=v[t]
\end{equation}

This relation holds up to and including the threshold check at the first-spike time. Thus:
\begin{equation}
L(a)
=
\inf\left\{
t\in\mathbb N_+:
a(1-\alpha^t)\geq\theta
\right\}
\end{equation}

When $a\leq0$, we have $v[t]\leq0<\theta$. When $0<a\leq\theta$, the inequality $0<1-\alpha^t<1$ gives:
\begin{equation}
v[t]=a(1-\alpha^t)<a\leq\theta
\end{equation}

Therefore, no finite first-spike time exists when $a\leq\theta$, and hence $L(a)=+\infty$.

When $a>\theta$, $v[t]$ increases monotonically and converges to $a>\theta$, so the first spike occurs in finite time. Since $\ln\alpha<0$, the threshold condition is equivalent to:
\begin{equation}
\begin{aligned}
a(1-\alpha^t)\geq\theta
&\iff \alpha^t\leq1-\frac{\theta}{a}\\
&\iff t\ln\alpha\leq\ln\left(1-\frac{\theta}{a}\right)\\
&\iff t\geq\frac{\ln(1-\theta/a)}{\ln\alpha}
\end{aligned}
\end{equation}
Taking the smallest positive integer satisfying this condition gives:
\begin{equation}
L(a)
=
\left\lceil
\frac{\ln(1-\theta/a)}{\ln\alpha}
\right\rceil, a>\theta
\end{equation}

\noindent\textbf{(2) Spike count within the time window.}

When $a\leq\theta$, the neuron never fires, so $C_{N_T}(a)=0$.

When $a>\theta$, the membrane potential resets after each spike to:
\begin{equation}
U[t]=U_{\mathrm{reset}}
\end{equation}

The constant input and fixed parameters ensure that the dynamics after each reset repeat the initial evolution. Consequently, consecutive spikes are separated by $L(a)$ time steps, and the $j$-th spike occurs at $jL(a)$. Thus:
\begin{equation}
\begin{aligned}
C_{N_T}(a)
&=
\#\left\{
j\in\mathbb N_+:
jL(a)\leq N_T
\right\}\\
&=
\left\lfloor
\frac{N_T}{L(a)}
\right\rfloor
\end{aligned}
\end{equation}
where $\#$ denotes the cardinality of the set.

\noindent\textbf{(3) Monotonicity.}

For any $a_1<a_2$ and $t\geq1$, the corresponding auxiliary sequences satisfy:
\begin{equation}
v_{a_2}[t]-v_{a_1}[t]
=
(a_2-a_1)(1-\alpha^t)>0
\end{equation}

If $L(a_1)<+\infty$, then at $t=L(a_1)$:
\begin{equation}
v_{a_2}[L(a_1)]
>
v_{a_1}[L(a_1)]
\geq\theta
\end{equation}

Therefore, $L(a_2)\leq L(a_1)$. If $L(a_1)=+\infty$, this inequality still holds in the extended real sense. Hence, $L(a)$ is non-increasing.

When $a_1>\theta$, the spike-count formula gives:
\begin{equation}
\begin{aligned}
C_{N_T}(a_2)
&=
\left\lfloor
\frac{N_T}{L(a_2)}
\right\rfloor\\
&\geq
\left\lfloor
\frac{N_T}{L(a_1)}
\right\rfloor\\
&=
C_{N_T}(a_1)
\end{aligned}
\end{equation}

When $a_1\leq\theta$:
\begin{equation}
C_{N_T}(a_1)=0\leq C_{N_T}(a_2)
\end{equation}

Thus, the spike count is non-decreasing in the input.
\subsection{Proof 2: Stability of competitive hash-bit decisions}
Fix a modality $m$ and a hash bit $k$, and define:
\begin{equation}
X_k^m[t] = Z_{k,+}^m[t]-Z_{k,-}^m[t] \in \{-1,0,1\}
\end{equation}
\begin{equation}
D_k^m = \sum_{t=1}^{N_T}X_k^m[t]
\end{equation}

The hash bit is determined by the following rule:
\begin{equation}
b_k^m =
\begin{cases}
+1,&D_k^m\geq0\\
-1,&D_k^m<0
\end{cases}
\end{equation}

Let $\widetilde X_k^m[t]$, $\widetilde D_k^m$, and $\widetilde b_k^m$ denote the corresponding perturbed quantities, with the same binary spike range and readout rule. If:
\begin{equation}
\#\left\{
t\in\{1,\ldots,N_T\}:
\widetilde X_k^m[t]\neq X_k^m[t]
\right\}
\leq r
\end{equation}
where $r\in\{0,\ldots,N_T\}$, then:
\begin{equation}
|D_k^m|>2r
\quad\Longrightarrow\quad
\widetilde b_k^m=b_k^m
\end{equation}

\textbf{Proof.} Let the set of time steps at which the spike difference changes be:
\begin{equation}
\mathcal A =
\left\{
t\in\{1,\ldots,N_T\}:
\widetilde X_k^m[t]\neq X_k^m[t]
\right\}
\end{equation}

By assumption, $|\mathcal A|\leq r$. Since the spike differences before and after perturbation both belong to $\{-1,0,1\}$, the magnitude of the change at each time step is at most $2$. Therefore:
\begin{equation}
\begin{aligned}
|\widetilde D_k^m-D_k^m|
&=
\left|
\sum_{t\in\mathcal A}
\bigl(\widetilde X_k^m[t]-X_k^m[t]\bigr)
\right|\\
&\leq
\sum_{t\in\mathcal A}
\left|
\widetilde X_k^m[t]-X_k^m[t]
\right|\\
&\leq2|\mathcal A|\\
&\leq2r
\end{aligned}
\end{equation}

If $|D_k^m|>2r$, there are two cases.

When $D_k^m>2r$:
\begin{equation}
\widetilde D_k^m \geq D_k^m-2r>0
\end{equation}

Hence:
\begin{equation}
\widetilde b_k^m=b_k^m=+1
\end{equation}

When $D_k^m<-2r$:
\begin{equation}
\widetilde D_k^m \leq D_k^m+2r<0
\end{equation}

Hence:
\begin{equation}
\widetilde b_k^m=b_k^m=-1
\end{equation}

The bit remains unchanged in both cases.

This proof uses only the bounded change in the spike difference at each time step and the upper bound on the number of affected time steps, allowing arbitrary dependencies across time steps. The condition $|D_k^m|>2r$ provides a sufficient condition for hash-bit stability.
\begin{figure*}[!ht]
    \centering
    \includegraphics[width=1\linewidth]{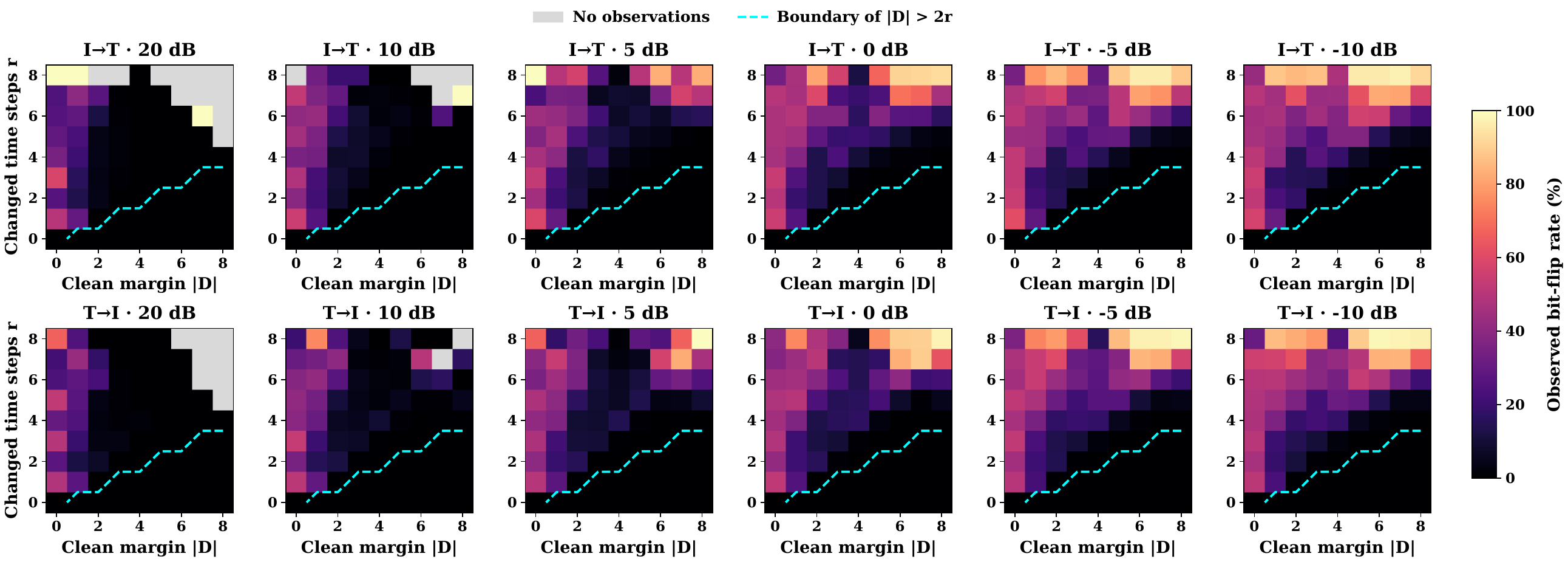} 
    \vspace{-20pt}
    \caption{Hash-bit flip rates under different noise levels and the stability guarantee of Proposition 2. The top and bottom rows correspond to image-to-text (I→T) and text-to-image (T→I) retrieval, respectively. The horizontal axis represents the absolute cumulative spike-count difference between the positive and negative channels under clean inputs ($\lvert D\rvert$), while the vertical axis represents the number of time steps at which the spike difference changes after perturbation ($r$). Colors indicate the observed bit-flip rate within each bin, and gray indicates bins with no observed samples. The cyan dashed line marks the boundary of the guaranteed stability region defined by $\lvert D\rvert>2r$. No bit flips are observed within this region, consistent with Proposition 2.}
\label{fig:flip_heatmaps}
\end{figure*}
\subsection{Empirical validation of competitive hash-bit stability}
Before feeding the features into the hashing network, we add perturbations to the offline-extracted CLIP query features (\(x\)) and compare the spike activity and hash bits before and after perturbation to evaluate the consistency of the observed readout behavior with Proposition 2.

Given a target signal-to-noise ratio ($s$) in dB, we first sample a standard Gaussian vector and normalize it to unit length:
\begin{equation}
    g\sim\mathcal{N}(0,I), u=\frac{g}{\lVert g\rVert_2}
    \end{equation}

We then scale the noise according to the L2 norm of the original features, obtaining the perturbed feature representation as follows:
\begin{equation}
    \widetilde{x}_s=x+10^{-s/20}\lVert x\rVert_2\,u
    \end{equation}

This setting ensures that the ratio of the noise L2 norm to the original feature L2 norm is $10^{-s/20}$. Thus, a lower signal-to-noise ratio corresponds to a stronger perturbation.

We use 16-bit hash codes and 8 time steps to evaluate six noise levels (20, 10, 5, 0, -5, and -10 dB) during inference. For I→T and T→I retrieval, we perturb only the image and text query features, respectively. 

Let $X_k[t]$ and $\widetilde X_k[t]$ denote the spike differences between the positive and negative channels before and after perturbation, respectively. For the $k$-th bit of each query, we compute the cumulative spike-count difference under clean input and the number of time steps at which the spike difference changes:
\begin{equation}
    D_k=\sum_{t=1}^{N_T}X_k[t],
r_k=\sum_{t=1}^{N_T}
\mathbf 1!\left[\widetilde X_k[t]\neq X_k[t]\right]
    \end{equation}

We then group the hash bits of all queries according to $(|D_k|, r_k)$  and compute the proportion of hash bits that differ before and after perturbation within each group.

As shown in \cref{fig:flip_heatmaps}, no bit flips occur in any observed group satisfying \(|D_k|>2r_k\) across both retrieval directions and all tested noise levels, consistent with Proposition 2. Even when noise alters spike firing at some time steps, the final hash bit remains unchanged as long as these changes are insufficient to reverse the cumulative spike-count advantage of one channel over the other.

Groups with zero or low bit-flip rates also exist outside the theoretically guaranteed region, reflecting the conservative nature of this sufficient condition. The bound \(2r_k\) assumes the maximum possible impact at each time step where the spike difference changes, whereas the actual changes may be smaller and may cancel each other across different time steps. Therefore, violating this condition does not necessarily lead to a bit flip.
\begin{figure*}[!ht]
    \centering
    \includegraphics[width=1\linewidth]{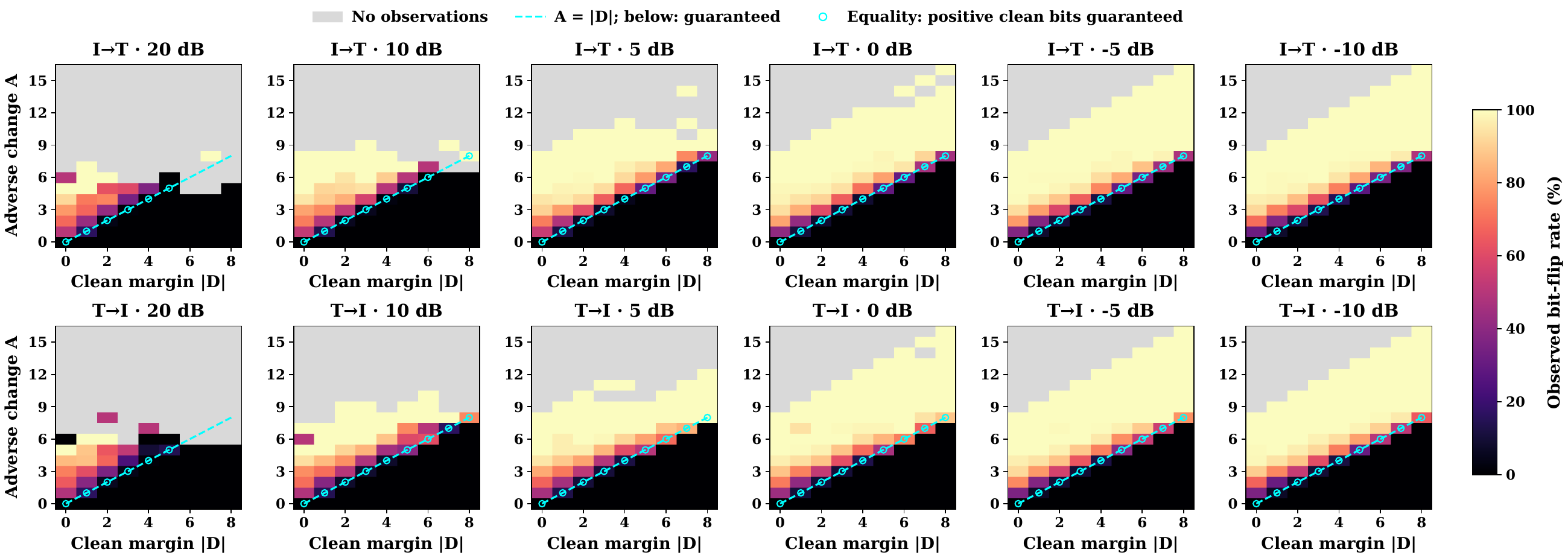} 
    \vspace{-20pt}
    \caption{Hash-bit stability analysis accounting for the direction and magnitude of spike changes. The top and bottom rows correspond to image-to-text (I→T) and text-to-image (T→I) retrieval, respectively, and each column corresponds to a different signal-to-noise ratio. The horizontal axis represents the absolute cumulative spike-count difference between the positive and negative channels under clean inputs, while the vertical axis represents \(A\), the sum of the magnitudes of changes that weaken the original bit decision. Colors indicate the observed bit-flip rate within each group, and gray indicates groups with no observed samples. The cyan dashed line marks the equality boundary between \(A\) and the absolute cumulative spike-count difference under clean inputs. All bits below this line satisfy the sufficient condition for stability. Circles mark observed groups on the equality boundary, where bits originally assigned \(+1\) remain guaranteed to be stable, whereas bits originally assigned \(-1\) are not guaranteed to be stable by this condition.}
\label{fig:directional_heatmap}
\end{figure*}
\subsection{Direction-aware stability conditions and coverage}
The preceding experiments show that some hash bits not covered by Proposition 2 remain stable. Counting only the number of time steps at which the spike difference changes cannot distinguish whether these changes weaken or strengthen the original bit decision. To address this limitation, we further consider the direction and magnitude of these changes to refine the stability condition and evaluate its coverage of actually stable hash bits.

Let $b$ denote the clean hash bit. The change in the spike difference at each time step is:
\begin{equation}
\Delta[t]=\widetilde X[t]-X[t]
\end{equation}

We define the total adverse change $A$ that weakens the original decision and the total favorable change $F$ that strengthens it:
\begin{equation}
\begin{aligned}
A&=\sum_{t=1}^{N_T}\max\!\left(-b\Delta[t],0\right),\\
F&=\sum_{t=1}^{N_T}\max\!\left(b\Delta[t],0\right)
\end{aligned}
\end{equation}

When the original bit is positive, $A$ accumulates the magnitudes of changes that decrease the count difference $D$; when the original bit is negative, $A$ accumulates the magnitudes of changes that increase $D$. In contrast, $F$ accumulates the magnitudes of changes that strengthen the original decision.

The perturbed cumulative difference therefore satisfies:
\begin{equation}
\begin{aligned}
b\widetilde D
&=bD+\sum_{t=1}^{N_T}b\Delta[t]\\
&=|D|+F-A\\
&\geq |D|-A.
\end{aligned}
\end{equation}

Thus, the following direction-aware condition is sufficient to guarantee that the hash bit remains unchanged:
\begin{equation}
\begin{cases}
D\geq A, & D\geq0\\[2mm]
|D|>A, & D<0
\end{cases}
\quad\Longrightarrow\quad
\widetilde b=b
\end{equation}

When $D\geq0$ and $D\geq A$, we have $\widetilde D\geq0$, so the perturbed hash bit remains positive. When $D<0$ and $|D|>A$, we have $\widetilde D<0$, so the perturbed hash bit remains negative. Both cases preserve the original bit decision. The different treatment of equality follows from the readout rule that assigns a positive bit when the cumulative difference is zero.

Since $A\leq2r$, every bit that satisfies the condition in Proposition~2 also satisfies the condition above. Proposition~2 requires only the number of changed time steps, whereas the direction-aware condition additionally uses the direction and magnitude of the changes, providing a guarantee at least as strong as the former. The latter remains a sufficient condition because it does not account for the compensation of adverse changes $A$ by favorable changes $F$.

We group all query hash bits by the absolute clean cumulative difference and the total adverse change. For a given noise level, we define:
\begin{equation}
\mathcal G_{d,a}
=
\left\{
(q,k):
|D_{qk}|=d,\;
A_{qk}=a
\right\}
\end{equation}
where $q$ and $k$ index queries and hash bits, respectively. The observed flip rate within each nonempty group is:
\begin{equation}
R_{\mathrm{flip}}(d,a)
=
\frac{
\displaystyle
\sum_{(q,k)\in\mathcal G_{d,a}}
\mathbf 1\!\left[\widetilde b_{qk}\neq b_{qk}\right]
}{
|\mathcal G_{d,a}|
},
|\mathcal G_{d,a}|>0
\end{equation}

\cref{fig:directional_heatmap} uses color to indicate this proportion and gray to indicate empty groups. The cyan dashed line corresponds to $A=|D|$. Below this line, $A<|D|$, which guarantees stability for both positive and negative bits.

Across both retrieval directions and all tested noise levels, no bit flips occur in the observed groups below the dashed line, consistent with the condition above. When $A$ exceeds the absolute clean cumulative spike-count difference between the positive and negative channels, many groups exhibit higher flip rates, with distinct regions of high flip rates under strong noise. This indicates that whether spike changes cause a bit flip depends not only on the number of changes but also on the cumulative extent to which they weaken the original decision.

On the boundary $A=|D|$, the guarantees differ for positive and negative bits. A positive bit retains its original decision even when the cumulative difference reaches zero, whereas a negative bit flips when the cumulative difference reaches zero. Consequently, groups that combine both bit polarities on the boundary can exhibit nonzero flip rates without contradicting the theory. Some groups above the dashed line also exhibit low flip rates because favorable changes $F$ can offset part of the adverse changes, allowing the final cumulative difference to retain the original decision.

\begin{figure}[!ht]
    \centering
    \includegraphics[width=1\linewidth]{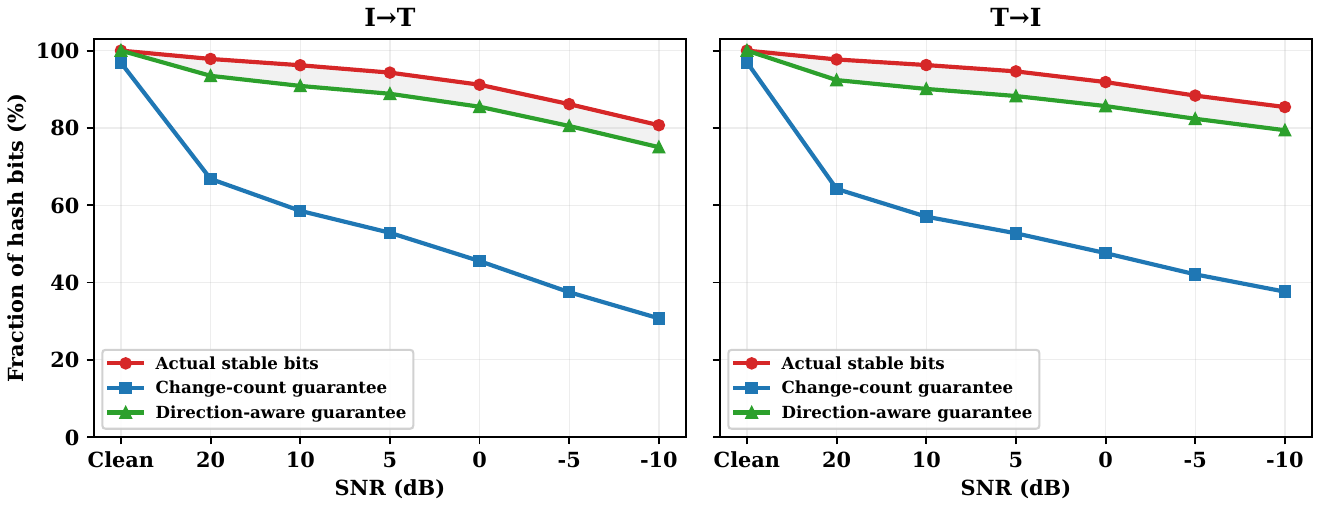} 
    \vspace{-20pt}
    \caption{Proportions of actually stable hash bits and guarantee coverage rates under two sufficient conditions at different noise levels. The red line represents the proportion of hash bits that remain unchanged, the blue line represents the proportion satisfying the change-count-based condition in Proposition 2, and the green line represents the proportion satisfying the direction-aware condition. All proportions are calculated relative to the total number of query hash bits.}
\label{fig:stability_coverage}
\end{figure}
The groups in the heatmaps contain different numbers of bits, so the area of the zero-flip region does not indicate the overall proportion of stable bits. \cref{fig:stability_coverage} therefore compares the observed stability rate with the guarantee coverage rates under the two conditions. Let $Q$ denote the number of queries, $K$ the code length, and $\mathbf 1[\cdot]$ the indicator function. The observed stability rate at signal-to-noise ratio $s$ is defined as:
\begin{equation}
S(s)
=
\frac{1}{QK}
\sum_{q=1}^{Q}\sum_{k=1}^{K}
\mathbf 1\!\left[
\widetilde b_{qk}(s)=b_{qk}
\right]
\end{equation}

The guarantee coverage rate under Proposition~2 is:
\begin{equation}
C_{\mathrm{count}}(s)
=
\frac{1}{QK}
\sum_{q=1}^{Q}\sum_{k=1}^{K}
\mathbf 1\!\left[
|D_{qk}|>2r_{qk}(s)
\right]
\end{equation}

For the direction-aware condition, we define the guarantee indicator for each bit as:
\begin{equation}
c_{\mathrm{dir}}(D,A)
=
\begin{cases}
\mathbf 1[D\geq A], & D\geq0\\
\mathbf 1[|D|>A], & D<0
\end{cases}
\end{equation}

The corresponding coverage rate is:
\begin{equation}
C_{\mathrm{dir}}(s)
=
\frac{1}{QK}
\sum_{q=1}^{Q}\sum_{k=1}^{K}
c_{\mathrm{dir}}\!\left(D_{qk},A_{qk}(s)\right)
\end{equation}

Here, $D_{qk}$ is determined by the clean input, while $r_{qk}(s)$ and $A_{qk}(s)$ denote the number of changed time steps and the total adverse change after perturbation, respectively. The inclusion relation between the conditions and their stability guarantees imply:
\begin{equation}
C_{\mathrm{count}}(s)
\leq C_{\mathrm{dir}}(s)
\leq S(s)
\end{equation}

\cref{fig:stability_coverage} shows that the guarantee coverage rates under both conditions decrease as the signal-to-noise ratio decreases. However, the direction-aware condition consistently covers more hash bits, and its coverage rate is closer to the observed stability rate. This suggests that estimating the effect of perturbations solely from the number of changed time steps can overestimate how much they weaken the original bit decision. Accounting for the direction and magnitude of the changes provides a more precise characterization of adverse changes and thus extends the coverage of the stability guarantee. A gap remains between the direction-aware coverage rate and the observed stability rate because this sufficient condition does not account for favorable changes that offset adverse changes, allowing some bits that do not satisfy the condition to remain stable.

\begin{figure*}[!ht]
    \centering
    \includegraphics[width=1\linewidth]{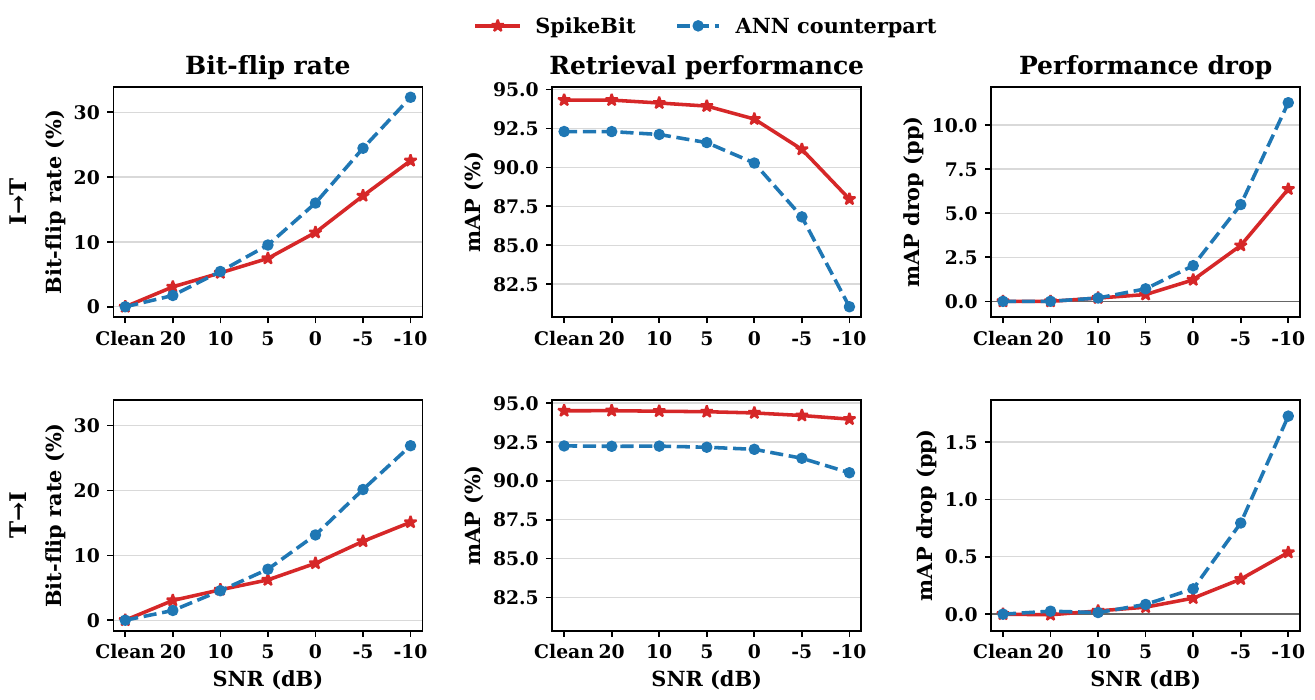} 
    \vspace{-20pt}
    \caption{Noise robustness of SpikeBit and the ANN counterpart. The figure compares hash-bit flip rates, mAP, and mAP drops relative to the noise-free results of each model. The horizontal axis represents the signal-to-noise ratio (SNR), and Clean indicates the absence of noise.}
\label{fig:snn_ann_snr}
\end{figure*}
\subsection{Effect of spiking dnamics on hash-bit stability}
Proposition 2 provides a sufficient condition for hash-bit stability based on the cumulative spike competition margin. Since continuous outputs also exhibit margin-based stability after sign quantization, we further compare SpikeBit with the ANN counterpart in \cref{tab:head_and_spiking_ablation} of the main paper to examine the contribution of spiking dynamics to encoding stability. Both models retain the same overall architecture and dual-channel competitive readout. They differ only in the neuron computation mechanism and receive identical perturbations to the query features.

As shown in \cref{fig:snn_ann_snr}, the two models exhibit similar bit-flip rates under weak perturbations. As the perturbation increases, SpikeBit shows lower bit-flip rates while maintaining higher retrieval accuracy and smaller performance drops. At -10 dB, the bit-flip rates of SpikeBit and the ANN counterpart are 22.52\% and 32.30\% for I→T retrieval, respectively. The corresponding mAP drops are 6.37 and 11.27 percentage points. For T→I retrieval, the bit-flip rates are 15.10\% and 26.90\%, respectively, while the mAP drops are 0.54 and 1.73 percentage points.

These results show that SpikeBit maintains greater hash-bit consistency under strong perturbations with the same dual-channel competitive structure. This advantage in encoding stability is also accompanied by smaller retrieval performance losses. These findings further demonstrate the contribution of spiking dynamics to encoding stability and retrieval robustness.
\begin{figure*}[!ht]
    \centering
    \includegraphics[width=1\linewidth]{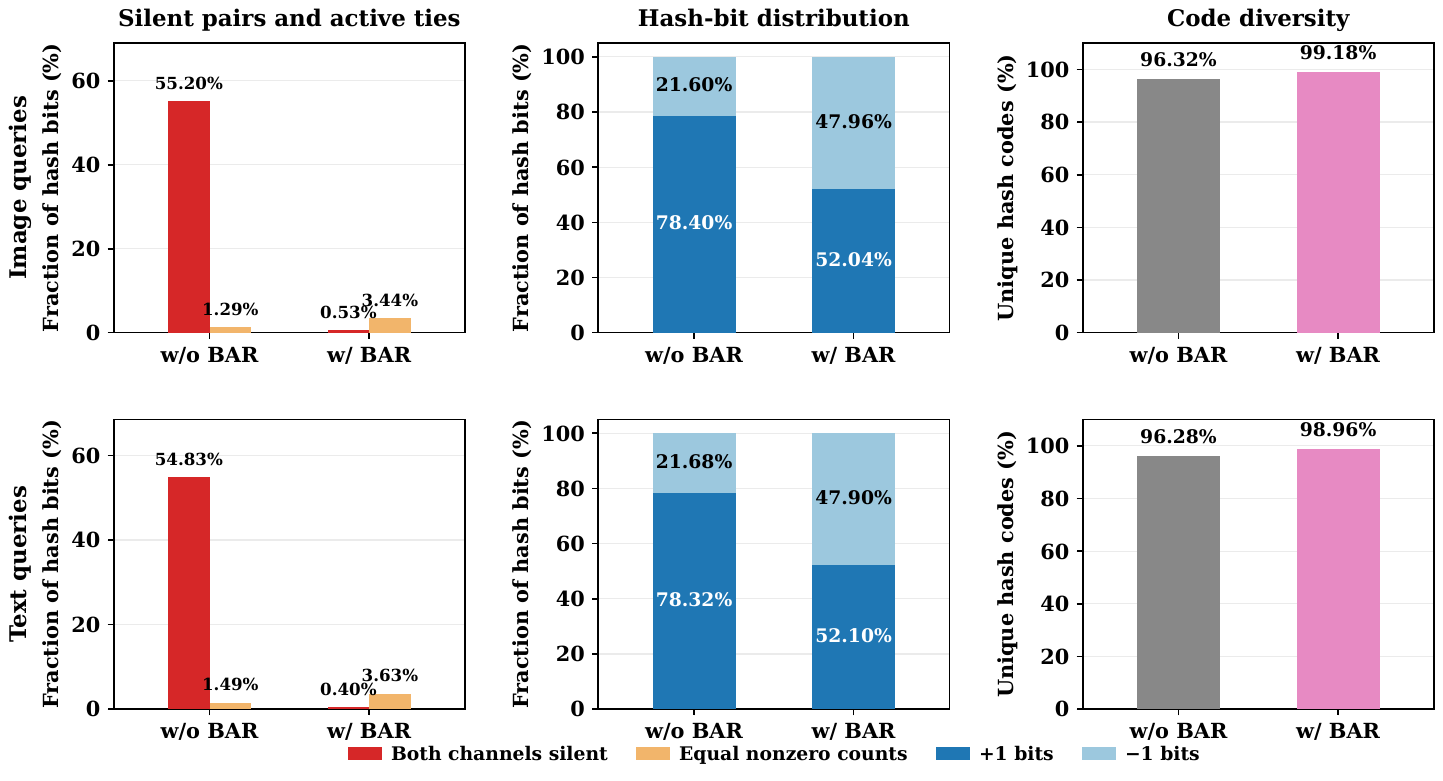} 
    \vspace{-20pt}
    \caption{Effect of BAR on channel silence and hash-code distribution. For each hash bit of each query, we record the cumulative spike counts of the positive and negative channels over the entire time window, denoted as $P$ and $N$, respectively. The dual-channel silence rate measures the proportion of bits with $P=N=0$. The proportion of bits with $P=N>0$ represents cases where both channels fire but have equal spike counts. We generate hash codes using the rule that assigns $+1$ when $P-N\geq0$ and $-1$ otherwise. We then calculate the proportions of positive and negative bits. The unique-code ratio is calculated by counting distinct hash codes across all queries and dividing by the total number of queries. A higher ratio indicates fewer duplicate codes.}
\label{fig:BAR_Effect}
\end{figure*}
\subsection{Effect of BAR on channel silence and hash-code distribution}
To examine whether BAR reduces default positive-bit decisions without spike support, we compare the hash head outputs with and without BAR. As shown in \cref{fig:BAR_Effect}, the dual-channel silence rates for image and text queries decrease from 55.20\% and 54.83\% to 0.53\% and 0.40\%, respectively. In contrast, the proportions of bits with equal nonzero spike counts increase from 1.29\% and 1.49\% to 3.44\% and 3.63\%, respectively. After introducing BAR, most remaining zero scores result from equal spike counts in the positive and negative channels. Dual-channel silence is substantially reduced, consistent with the design objective of BAR.

As silent bits decrease, the proportions of positive bits for image and text queries decrease from 78.40\% and 78.32\% to 52.04\% and 52.10\%, respectively. The overall distribution of positive and negative bits becomes more balanced. The unique-code ratios increase from 96.32\% and 96.28\% to 99.18\% and 98.96\%, respectively, indicating fewer duplicate hash codes. These results explain the performance gains reported in \cref{tab:loss_ablation} of the main paper. They also support the role of BAR in reducing default positive-bit decisions by suppressing dual-channel silence and improving hash-code distribution and diversity.

\begin{figure*}[!ht]
    \centering
    \includegraphics[width=1\linewidth]{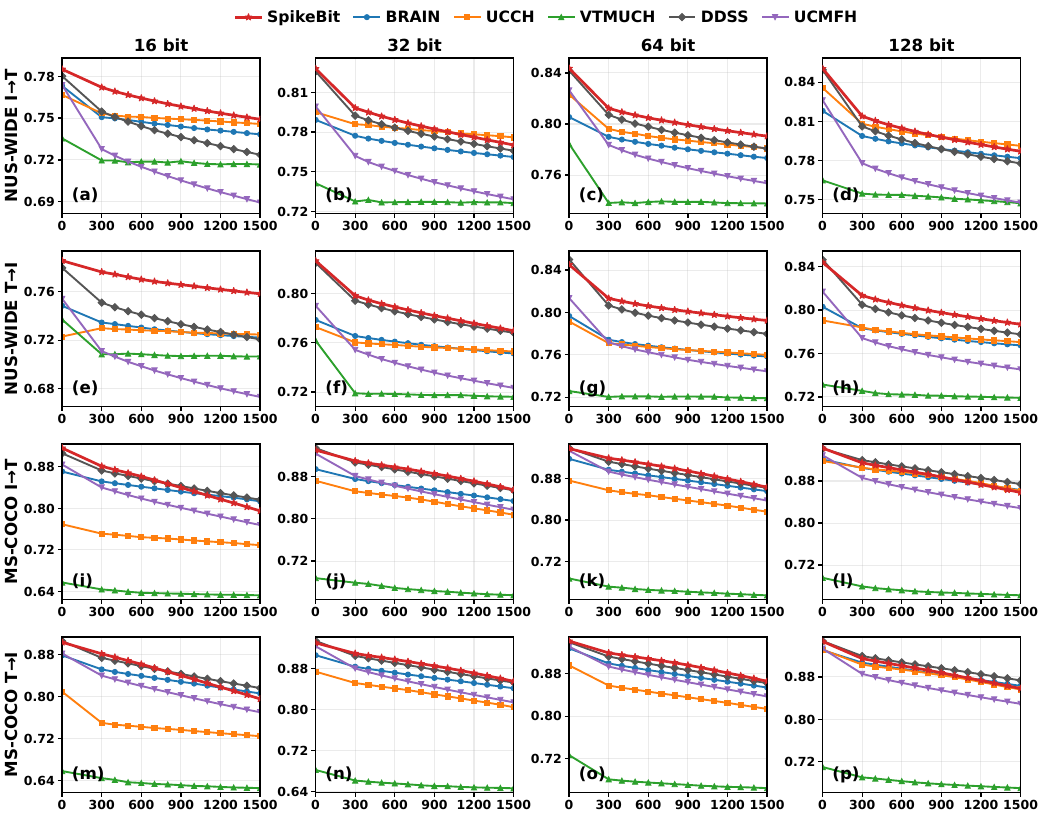} 
    \vspace{-20pt}
    \caption{Top-(N) precision curves on NUS-WIDE and MS-COCO. The horizontal axis shows the number of retrieved results (N), and the vertical axis shows Precision@(N) averaged across all queries.}
\label{fig:coco_topk}
\end{figure*}
\section{Top-$N$ precision comparison}\label{topn}
\cref{fig:coco_topk} compares bidirectional Top-N retrieval precision across four code lengths on NUSWIDE and MSCOCO. On NUSWIDE, SpikeBit consistently leads in both retrieval directions across the entire displayed range with 16-bit codes, indicating that compact binary representations effectively preserve semantic information for cross-modal ranking. With 64-bit codes, SpikeBit consistently leads in image-to-text retrieval and achieves the best precision at nearly all cutoffs in text-to-image retrieval. The 32-bit and 128-bit codes also remain competitive, particularly in text-to-image retrieval, where the advantage extends to larger retrieval cutoffs.

On MSCOCO, SpikeBit achieves the best or near-best precision across a broad range of cutoffs in both retrieval directions with 32-bit and 64-bit codes. It also remains competitive at the other code lengths. Compared with BRAIN, which also employs a spiking mechanism, SpikeBit achieves higher precision in most settings, supporting the effectiveness of the overall encoding design. 

Although DDSS performs better in some settings, SpikeBit achieves comparable or better precision in most settings. Meanwhile, SpikeBit requires fewer parameters and operations, consumes less estimated energy, and achieves approximately 19x faster encoding than DDSS. Together with the computational cost analysis presented in \cref{fig:efficiency_1x4} of the main paper, these results demonstrate that the proposed lightweight design balances cross-modal ranking quality and encoding efficiency.
\begin{figure*}[!ht]
    \centering
    \includegraphics[width=1\linewidth]{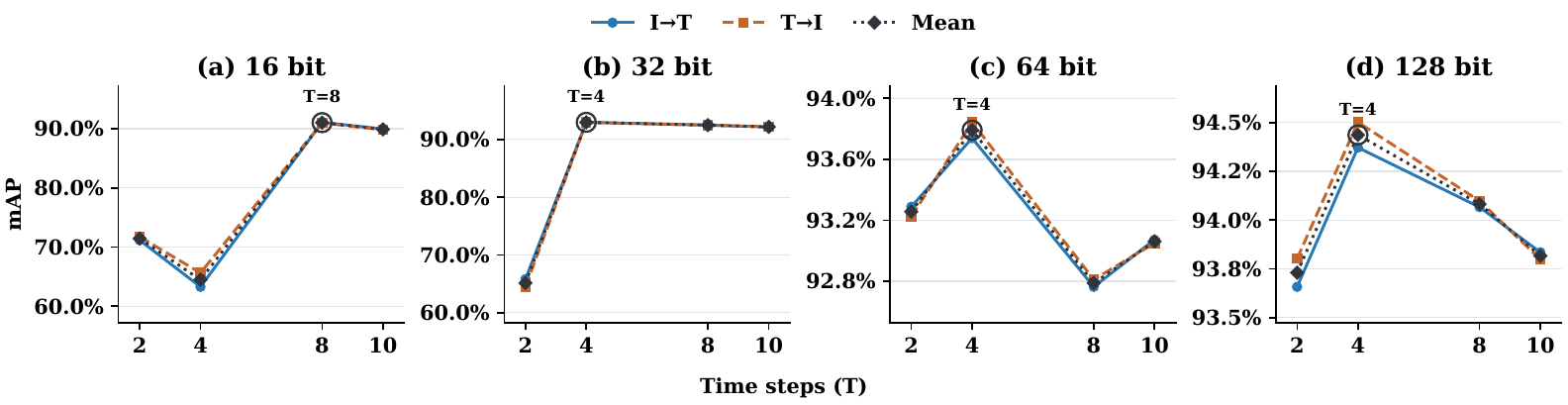} 
    \vspace{-20pt}
    \caption{Effect of the spike time window length on retrieval performance across different code lengths. The blue and orange lines represent the mAP for image-to-text and text-to-image retrieval, respectively, while the black line represents the bidirectional average. Circles mark the selected time window length.}
\label{fig:time_step_analysis}
\end{figure*}
\begin{figure*}[!ht]
    \centering
    \includegraphics[width=1\linewidth]{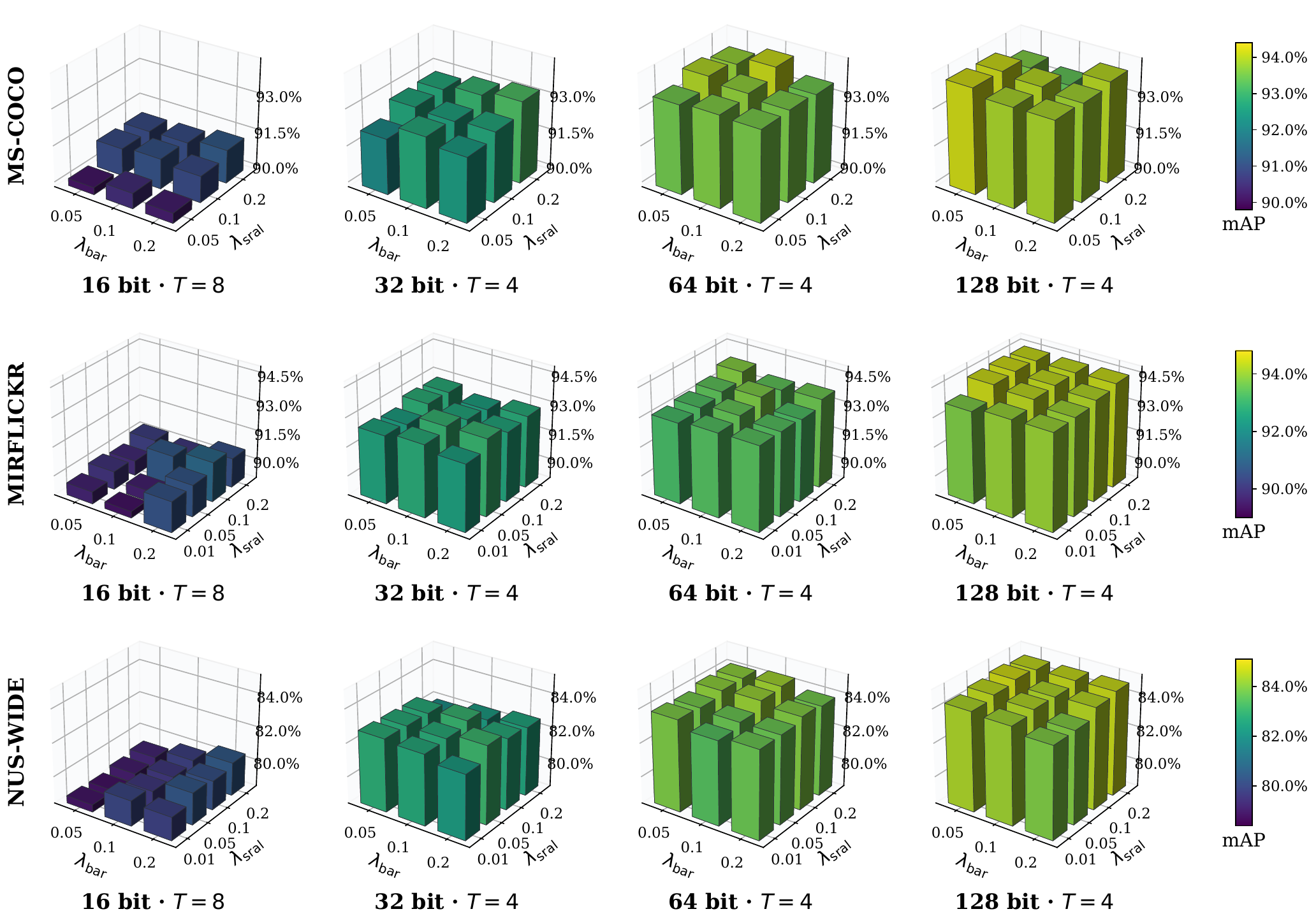} 
    \vspace{-20pt}
    \caption{Retrieval performance of SpikeBit under different loss weights. Each bar represents a combination of loss weights $\lambda_{\mathrm{bar}}$ and $\lambda_{\mathrm{sral}}$. The bar height and color indicate the average of the best mAP values for image-to-text and text-to-image retrieval. We use $T=8$ for 16-bit codes and $T=4$ for the other code lengths.}
\label{fig:loss_weights}
\end{figure*}
\section{Time window length}\label{timewind}
\cref{fig:time_step_analysis} compares retrieval performance across four code lengths under different time windows. The 16-bit codes achieve the highest average precision across both retrieval directions at T=8, while the 32-bit, 64-bit, and 128-bit codes perform best at T=4. These results indicate that the time window should be selected according to the code length. Compact codes may require longer spike sequences to form discriminative count differences, whereas longer codes achieve good performance within shorter windows. However, increasing the number of time steps does not necessarily improve retrieval, and extending the window beyond the selected length provides no further gains at any code length. Based on performance in both retrieval directions, we set T=8 for 16-bit codes and T=4 for 32-bit, 64-bit, and 128-bit codes.
\section{Loss weights}\label{lossweigh}
We fix the time window length for each setting and jointly sweep $\lambda_{\mathrm{bar}}$ and $\lambda_{\mathrm{sral}}$ to compare retrieval performance across weight combinations. \cref{fig:loss_weights} shows that the degree of performance variation differs across datasets and code lengths. On MIRFLICKR, we average the best mAP values from the two retrieval directions. Across the tested combinations, the difference between the highest and lowest average mAP values is 0.0171 for 16-bit codes. The corresponding differences for 32-bit, 64-bit, and 128-bit codes range from 0.0058 to 0.0072. Some longer codes also achieve similar mAP values under neighboring weight combinations.

These results indicate greater performance variation across weight combinations in the 16-bit setting. One possible reason is the limited representation capacity of short codes. The balance between the two losses may therefore have a greater influence on the cross-modal semantic information retained in a limited number of bits, thereby affecting retrieval rankings. For each dataset and code length, we select the weight combination that achieves the highest average mAP.

\section{Qualitative comparison}\label{quacompa}
To qualitatively compare the retrieval results of SpikeBit and the baselines, we present the Top-10 results of each method on challenging cases.
\subsection{Successful retrieval cases}
\begin{figure*}[!ht]
    \centering
    \includegraphics[width=0.9\linewidth]{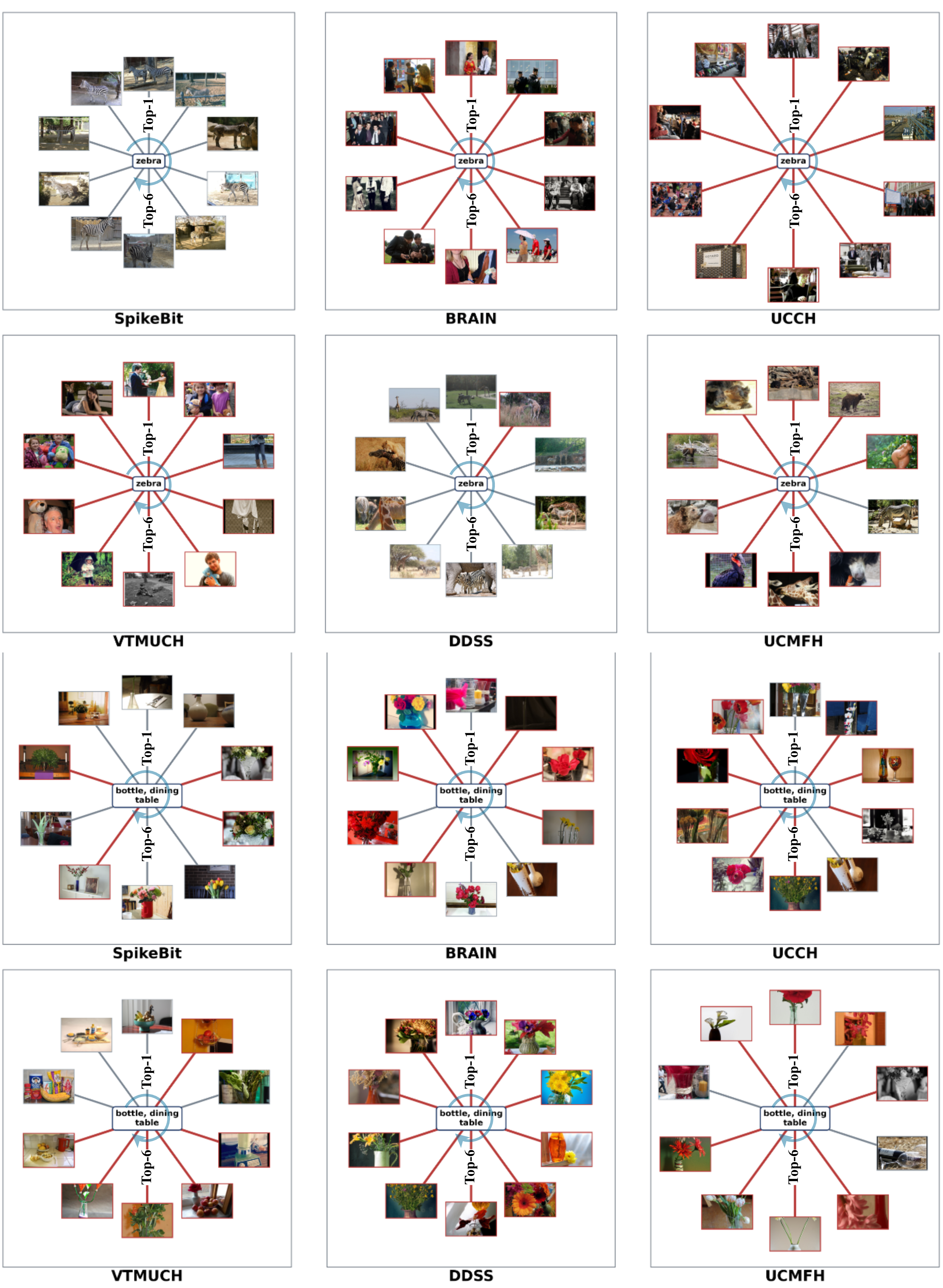} 
    \vspace{-10pt}
    \caption{Text-to-image retrieval for the challenging samples \textit{zebra} and \textit{bottle, dining table}. The Top-10 results of SpikeBit and the baselines are compared under the same conditions. Edges represent Hamming distances. Gray and red edges indicate whether the retrieved images share categories with the original paired image, respectively.}
\label{fig:positive_1}
\end{figure*}

As shown in \cref{fig:positive_1}, SpikeBit retrieves relevant images at all Top-10 positions for the \textit{zebra} query, covering different viewpoints and backgrounds. DDSS also retrieves several relevant images, while some baselines repeatedly return images of people or other animals. For the multi-label query \textit{bottle, dining table}, SpikeBit ranks several images sharing at least one query category among the top results.

\begin{figure*}[!ht]
    \centering
    \includegraphics[width=0.9\linewidth]{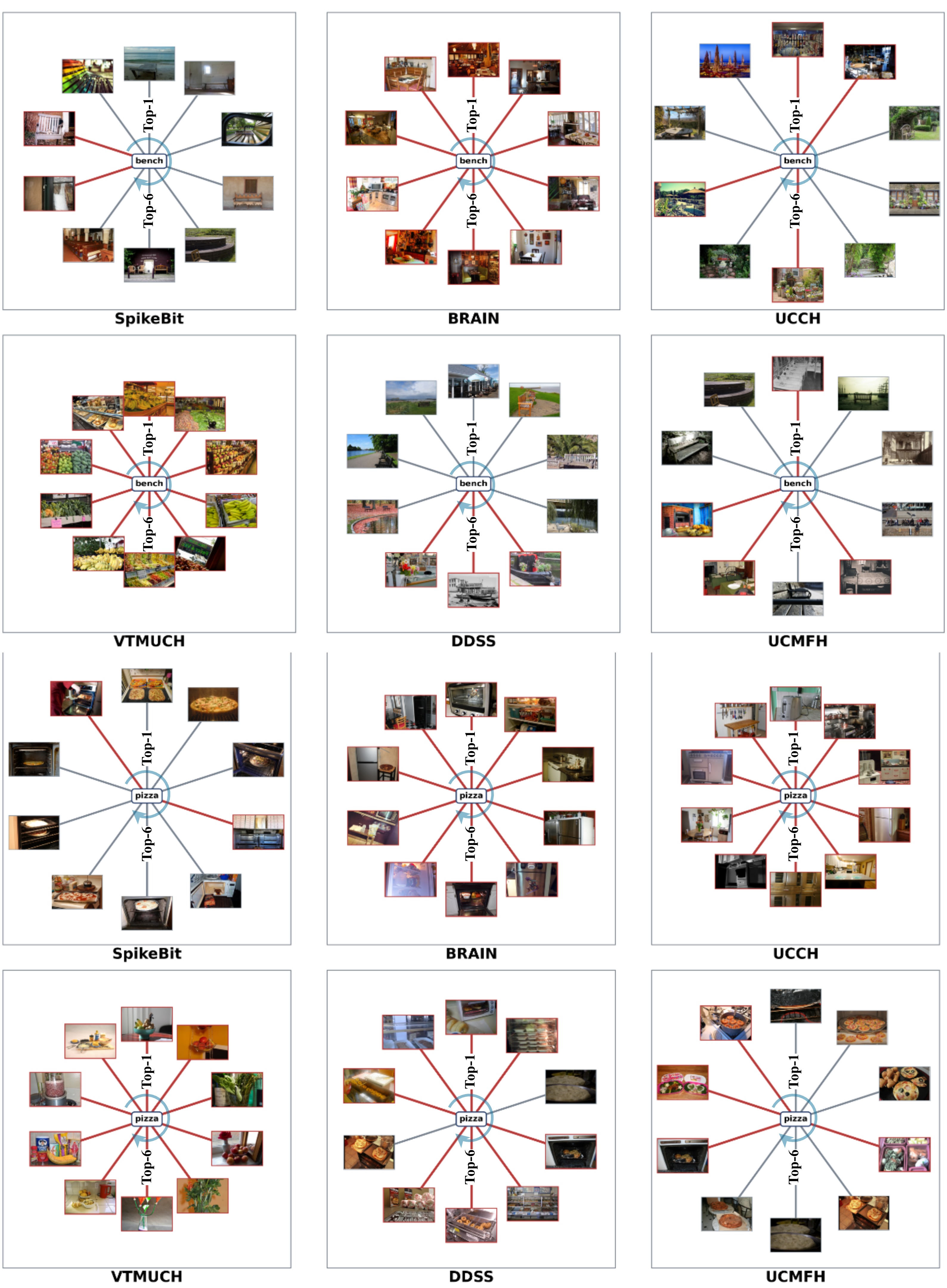} 
    \vspace{-10pt}
    \caption{Text-to-image retrieval for the challenging samples \textit{bench} and \textit{pizza}. The Top-10 results of SpikeBit and the baselines are compared under the same conditions. Edges represent Hamming distances. Gray and red edges indicate whether the retrieved images share categories with the original paired image, respectively.}
\label{fig:positive_2}
\end{figure*}

In \cref{fig:positive_2}, SpikeBit ranks benches in seaside, indoor, and outdoor scenes among the top results for the \textit{bench} query. DDSS also retrieves several relevant images. The \textit{pizza} case further illustrates the distinction between target objects and co-occurring scenes. Several methods retrieve images of ovens or kitchen scenes. However, SpikeBit retrieves more images labeled pizza in the Top-10 than DDSS. These results demonstrate the stability of our method in retrieving challenging samples.
\subsection{Failed retrieval cases}
\begin{figure*}[!ht]
    \centering
    \includegraphics[width=0.9\linewidth]{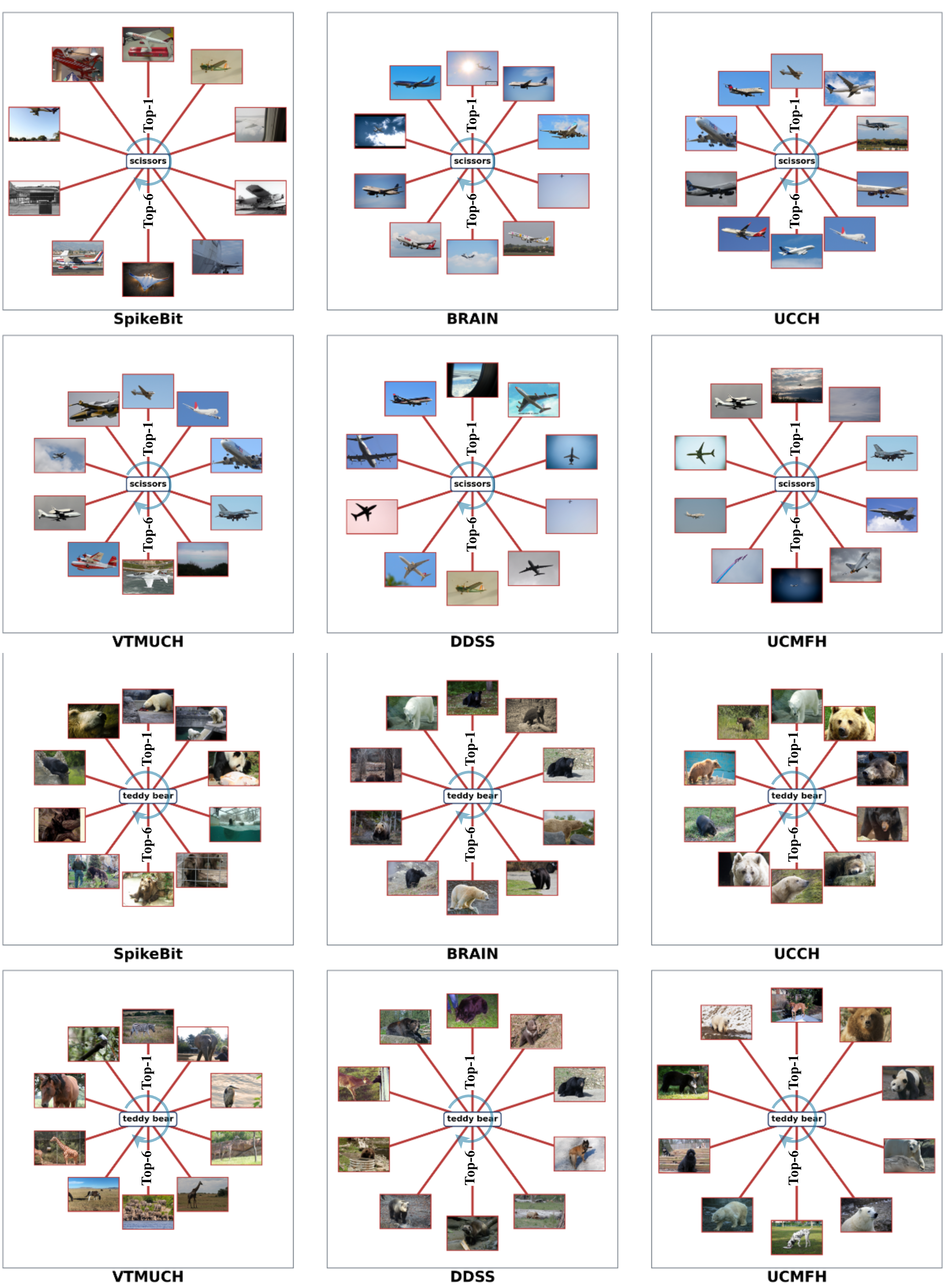} 
    \vspace{-10pt}
    \caption{Text-to-image retrieval for the challenging samples \textit{scissors} and \textit{teddy bear}. The Top-10 results of SpikeBit and the baselines are compared under the same conditions. Edges represent Hamming distances. Gray and red edges indicate whether the retrieved images share categories with the original paired image, respectively.}
\label{fig:negetive_1.png}
\end{figure*}

As shown in \cref{fig:negetive_1.png}, both failure cases show that incorrect categories dominate the top-ranked results. For the \textit{scissors} query, almost all retrieved images depict airplanes, and none of the six methods retrieves a relevant image. This indicates that the hash-based ranking places visually similar airplane images ahead of truly relevant candidates. Such results reveal a common retrieval bias.

The incorrect results for the \textit{teddy bear} query exhibit a clearer semantic pattern. SpikeBit and several baselines mainly retrieve images of real bears, pandas, or other animals. Teddy bears and real animals share some naming and visual characteristics. However, distinguishing them requires recognizing the attribute of being a toy. These results indicate that general animal similarity can produce a consistent group of nearest neighbors but still fails to satisfy fine-grained category requirements.
\end{document}